\documentclass[
 aip,
 amsmath, 
 amssymb,
 reprint,
onecolumn 
]{revtex4-1}
\usepackage{graphicx}
\usepackage{dcolumn}
\usepackage{bm}
\usepackage[utf8]{inputenc}
\usepackage[T1]{fontenc}
\usepackage{etoolbox}
\usepackage{booktabs} 
\usepackage{array} 
\usepackage{multirow} 
\usepackage{makecell} 
\usepackage{threeparttable}
\usepackage{tabularx}
\usepackage{float} 
\usepackage{url}

\begin{document}
\preprint{AIP/123-QED}
\title{Discovery of Green's function based on symbolic regression with physical hard constraints}
\author{Jianghang Gu}
\affiliation{College of Engineering, Peking University, Beijing 100871, P. R. China.}
\author{Mengge Du}
\affiliation{College of Engineering, Peking University, Beijing 100871, P. R. China.}
\author{Yuntian Chen}
\email{ychen@eitech.edu.cn}
\affiliation{Ningbo Institute of Digital Twin, Eastern Institute of Technology, Ningbo, Zhejiang 315200, P. R. China}
\author{Shiyi Chen}
\email{syc@coe.pku.edu.cn, chensy@sustech.edu.cn}
\affiliation{College of Engineering, Peking University, Beijing 100871, P. R. China.}

\begin{abstract}
\textbf{Abstract}\\
The Green's function, serving as a kernel function that delineates the interaction relationships of physical quantities within a field, holds significant research implications across various disciplines. It forms the foundational basis for the renowned Biot-Savart formula in fluid dynamics, the theoretical solution of the pressure Poisson equation, and the closure modeling of turbulent Reynolds stresses. Despite their importance, the theoretical derivation of the Green's function is both time-consuming and labor-intensive.
In this study, we employed DISCOVER, an advanced symbolic regression method leveraging symbolic binary trees and reinforcement learning, to identify unknown Green's functions for several elementary partial differential operators, including Laplace operators, Helmholtz operators, and second-order differential operators with jump conditions. The Laplace and Helmholtz operators are particularly vital for resolving the pressure Poisson equation, while second-order differential operators with jump conditions are essential for analyzing multiphase flows and shock waves.
By incorporating physical hard constraints, specifically symmetry properties inherent to these self-adjoint operators, we significantly enhanced the performance of the DISCOVER framework, potentially doubling its efficacy. Notably, the Green's functions discovered for the Laplace and Helmholtz operators precisely matched the true Green's functions. Furthermore, for operators without known exact Green's functions, such as the periodic Helmholtz operator and second-order differential operators with jump conditions, we identified potential Green's functions with solution error on the order of $10^{-10}$. 
This application of symbolic regression to the discovery of Green's functions represents a pivotal advancement in leveraging artificial intelligence to accelerate scientific discoveries, particularly in fluid dynamics and related fields. Our findings underscore the promise of this approach for further exploration and its potential to transform the understanding and modeling of complex physical systems.
Code and data is available at \url{https://github.com/hangjianggu/Discover_Green_function/tree/main}.
\end{abstract}

\maketitle
Green's functions, integral to both mathematics and physics, represent a pivotal concept employed in solving inhomogeneous differential equations under specific initial or boundary conditions \cite{stakgold2011green}. Functionally, they serve as the impulse response of a given system, elucidating the system's reaction to a localized input. By convolving the Green's function with the source term—depicting the localized input, this technique facilitates the solution of differential equations, streamlining the process significantly.
In the field of fluid dynamics, the Green's function serves as the foundation for the well-known Biot-Savart Law formula, which describes the mapping from vorticity fields to velocity fields \cite{wu2015vortical}. The Green's function is also central to the theoretical solution of the pressure Poisson equation, which is crucial for solving the Navier-Stokes flow field, and even for solving the turbulent fluctuating velocity equations and modeling the closure of turbulent Reynolds stresses. For example, Kraichnan \cite{kraichnan1987eddy} utilized the Green's function to derive an exact nonlinear algebraic stress formulation from the fluctuating velocity equations. This work was further developed and refined by Hamba \cite{hamba2005nonlocal}, culminating in the representation of the Reynolds stress as the global integral of the product between the unit eddy viscosity coefficient tensor and the mean velocity gradient.
Unfortunately, mathematically deducing Green's functions can indeed pose challenges owing to the intricacy of the equations and the specific conditions they necessitate. 
Nowadays, with the advent of artificial intelligence (AI), this laborious process has experienced a significant acceleration. 

AI has been successfully used in many scientific and engineering fields. 
Since the Dartmouth Conference in 1956, the development of AI has gone through two stages: symbolic AI, represented by expert systems; and connectionist AI, represented by data-driven models \cite{zhang2023toward, chen2021theory}. 
Connectionist AI encompasses many data-driven machine learning models and is the current mainstream approach in deep learning, having been successfully applied in numerous fields \cite{milano2002neural, meng2020composite}. 
For example, the CNN and UNet \cite{guo2016convolutional} parameterize the function map between input and output base on multiple convolution kernels.
the Fourier neural operator \cite{li2023fourier} aims to parameterize the integral kernel in the Fourier space. 
DeepONet \cite{lu2021learning} extends the universal approximation theorem for operators \cite{chen1995universal} to deep neural networks.
These methods attempt to discover the coefficients of PDE models or learn the operators that map excitation to system responses.
However, these connectionist AI methods depend heavily on training data, while obtaining data is always expensive and time-consuming in actual scientific and engineering scenarios. Their essence is surrogate models (i.e., unexplainable black boxes), and no explicit knowledge can be obtained.

Symbolic AI utilizes domain knowledge through expert systems \cite{laporte1989design, liao2005expert} and knowledge engineering \cite{studer1998knowledge} to develop AI systems. The mainstream method of symbolic AI is knowledge discovery methods, which are capable of directly extracting the governing equation that best matches the data with transfer ability when the equation structure is unknown. The core of these methods is to construct candidate sets (i.e., library of function terms) based on prior knowledge, and then choose the most appropriate combination of candidate terms to produce equations through various optimization methods. Based on candidate terms, existing methods can be divided into closed library methods, expandable library methods, and open-form equation methods \cite{chen2022integration}. 

Closed library methods are the most widely used, and are based on sparse regression for the distillation of the dominating candidate function terms.
For example, Brunton et al. \cite{brunton2016discovering} developed SINDy, which built an overcomplete library first and then utilized sparse regression to determine the appropriate equation. PDE-FIND extends SINDy to PDEs by introducing partial derivative terms in the library \cite{rudy2017data}.
SGTR combines group sparse coding and solves the problem of parametric PDEs \cite{rudy2019data}. Closed library methods can identify most of the governing equations of simple systems, but it is difficult to construct an overcomplete candidate set for complex systems. 

Compared with closed libraries, expandable libraries are more suitable for discovering governing equations with complex structures. DLGA integrates the neural network and the genetic algorithm, and realizes automatic expansion of the candidate set by encoding different function terms as gene segments \cite{xu2020dlga}. In addition to genetic algorithms, PDE-Net 2.0 \cite{long2019pde} introduces SymNet \cite{sahoo2018learning}, which uses network topology to generate interaction terms. Nevertheless, both PDE-Net 2.0 and genetic algorithms can only generate new function terms through addition and multiplication, and cannot implement division operations or generate composite functions. As a consequence, despite the fact that expandable library methods are more flexible and use fewer resources than closed library methods \cite{long2019pde}, they are still unable to construct governing equations with fractional structures and compound functions. 

In order to extract arbitrary equations from data, open-form equation methods are proposed. For instance, automated reverse engineering automatically generates equations for a nonlinear coupled dynamical system with the assistance of symbolic mathematics \cite{bongard2007automated}. However, because this method examines each variable separately, there are scalability issues. Later, researchers conducted more work on symbolic regression, and recommended that the governing equation be represented by binary trees \cite{schmidt2009distilling}. SGA provides a tree-based genetic algorithm that can handle increasingly complex systems and accomplish PDE discovery using symbolic mathematical representation \cite{chen2022symbolic}. 

Due to the wider optimization space of open-form equation methods, they have greater computational cost than conventional methods in practice. 
To ameliorate the limitations of fixed candidate libraries and accelerate the search process, DISCOVER is proposed by Du et al.\cite{du2024discover, du2024physics}. A strengthened version of DISCOVER based on Large Language Models is also proposed by Du et al. \cite{du2024llm4ed} recently. Compared to other symbolic frameworks, DISCOVER enables the efficient discovery of quantitatively accurate expressions from complex nonlinear systems while including the minimal possible number of function terms. 
Enhanced by structural information and reinforcement-learning optimization, DISCOVER demonstrates superior performance compared to other well-known symbolic regression methods such as PySR \cite{cranmer2023pysr}.

However, it is still challenging to derive a concise and logical Green's function solely using DISCOVER due to the vast number of random combinations. DISCOVER, by relying solely on the combinations of symbols without any prior knowledge, results in an extremely large search space. Consequently, it requires more search time and may become trapped in local optima, failing to find the optimal solution.
Integrating domain knowledge and prior information into DISCOVER to achieve a dual-driven model of knowledge and data can fundamentally improve its performance \cite{xu2024worth}. Therefore, in this paper, we strive to discern the Green's function of the governing PDEs from sparse observed data using DISCOVER with physical hard constraints. Specifically, the physical hard constraint is constructed based on the symmetry property of Green's function. Since the governing equation can be presented by a expression tree via symbolic mathematics, it can be further split into left and right sub-trees, with a transformation of symmetrical design.
From an optimization perspective, hard constraints are more efficient methods than soft constraints, in general. Hard constraints can ensure that a given governing equation is strictly satisfied over a certain region or the entire domain. Theoretically, the application of hard constraints enables the model to require fewer data, achieve higher prediction accuracy, and obtain stronger robustness to noisy observations \cite{chen2021theory}. 
To show the effectiveness of hard constraints, ablation studies are given in section \ref{sec3}.

It should be noted that there also exists a few works which compute Green’s functions to solve PDEs based on deep learning methods. Specifically, a rational neural network structure \cite{boulle2022data} was proposed to train networks with generated excitation for approximating Green’s functions. 
For handling nonlinear boundary value problems, DeepGreen \cite{gin2021deepgreen} was put forward using the autoencoder structure. 
Teng et al. \cite{teng2022learning} proposed a novel neural network method, “GF-Net”, for learning the Green’s functions of the classic linear reaction-diffusion equation in the unsupervised fashion. 
While excellent for learning Green's functions, these methods do not yield an explicit expression for the Green's function, and the correctness of the Green's function obtained through neural network training remains to be further examined. 
They are essentially black-box models, capable only of mimicking the mapping relationships of Green's functions without providing results in the form of equations. Due to their black-box nature, it is challenging for researchers to further analyze the theoretical significance of Green's functions based on these models' results, thus limiting deeper exploration.
Therefore, in this article, we propose the adoption of DISCOVER with physical hard constraints to deduce the unknown expression of Green's function. 
To the best of our knowledge, this is the first time the explicit expression of Green's function has been found by means of deep learning methods. 
During the validation phase, we conducted comparative analysis for PDE operators (i.e., Laplace operator, Helmholtz operator) with ground-truth Green's function expressions. For those without (i.e., Periodic helmholtz operator, second-order differential operator with jump conditions), we undertook a series of rationality analyses on our results. Our contributions are summarized as follows:
\begin{itemize}
    \item The introduction of a universal method for uncovering latent Green's functions. These discovered Green's functions provide insights into the physical properties of the operator $\mathcal{L}$ and the types of boundary constraints imposed.
    \item Utilization of the discovered Green's functions enables the direct acquisition of solutions $u$ for new forcing terms $f$, eliminating the need for a retraining process.
    \item For the first time, AI has been used to automatically extract explicit formula of Green's functions from observational data, leading to the discovery of the Green's functions for the periodic Helmholtz operator and the second-order differential operator with jump conditions.
    \item Achieve a solution error up to the order of $10^{-10}$. Provides opportunities for further research into theoretical solutions of the pressure Poisson equation in fluid mechanics and issues such as multiphase flow.
\end{itemize}

This paper comprises four sections. In section \ref{sec2}, the physical hard constraint strategy based on DISCOVER is introduced. In section \ref{sec3}, numerical results and validations are demonstrated. Finally, conclusions and outlooks are given in section 4.

\section{Methodology\label{sec2}}

\begin{figure}[H]
    \centering
    \includegraphics[width=1\textwidth]{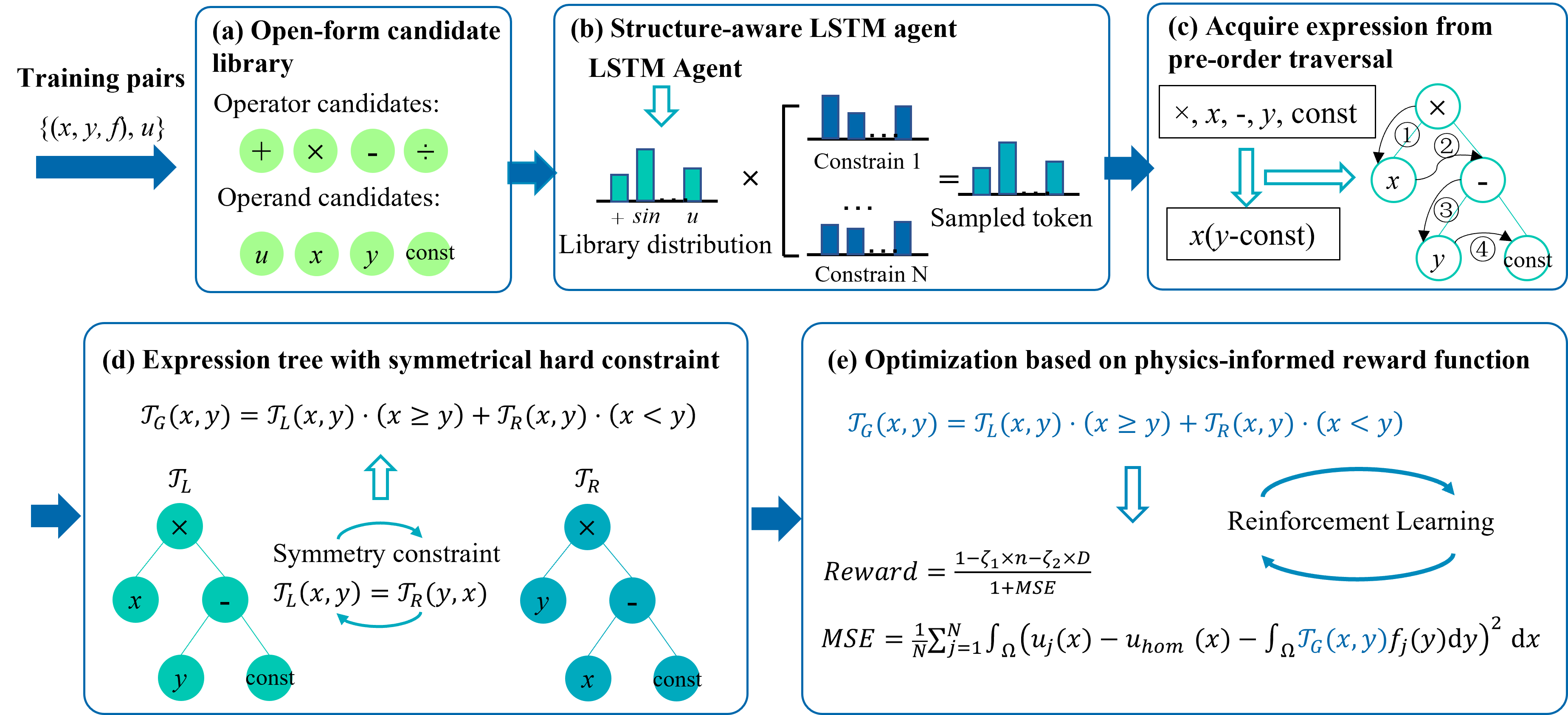}
        \caption{Overview of the mining of Green's function based on DISCOVER framework \cite{du2024discover}. (a) Establishing open-form candidate library. (b) Structure-aware LSTM agent for generating possible expressions with imposing constraints. (c) Acquire expression from pre-order traversal. (d) Rebuild the symbolic tree based on symmetrical hard constraint. (e) Reward function based on physics-informed constraints in Eq. (\ref{eq6}).}
    \label{fig1}
\end{figure}

\paragraph*{\textbf{Overall framework.}}
In this paper, we propose a novel and universal framework to discover unknown Green's functions from observational data, effectively bypassing the laborious processes of mathematical deduction. 
The overall framework can be illustrated in FIG. \ref{fig1}.
Specifically, we leverage symbolic regression method to unveil the Green's function of linear differential equations $\mathcal{L} u=f$ from input-output pairs $(f, u)$, rather than directly learning $\mathcal{L}$ or its parameters. 
The training dataset is first established using a Gaussian process. 
As for symbolic regression, we employ the DISCOVER framework \cite{du2024discover} to realize AI-based scientific discovery. 
Based on the DISCOVER framework, we design symmetrical hard constraints for the generated symbolic trees. Inspired by PINN framework, a reward function is then designed to find the unknown Green's function. Details for each step is illustrated in following subsections.

\subsection{Preparation for finding Green’s functions.}

\begin{figure}[H]
    \centering
    \includegraphics[width=0.85\textwidth]{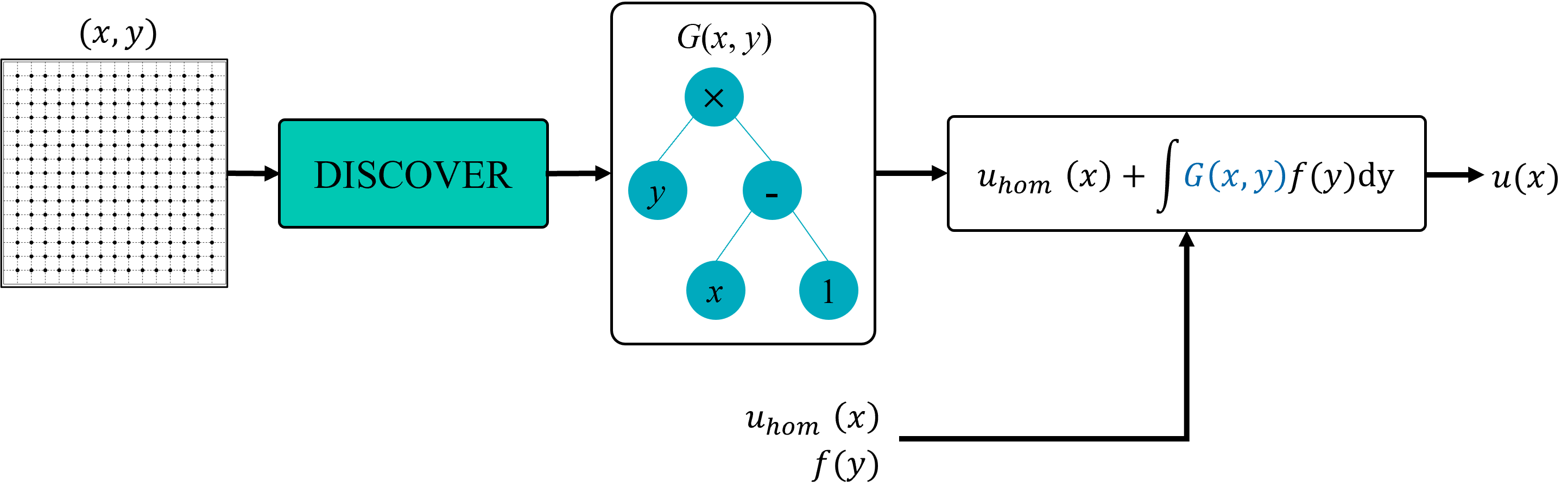}
        \caption{Schematic of learning Green's function based on DISCOVER \cite{du2024discover}.}
    \label{fig11}
\end{figure}

\paragraph*{\textbf{Learning Green’s functions.}}
We consider linear differential operators, $\mathcal{L}$, defined on a bounded domain $\Omega \subset \mathbb{R}^d$, where, $d \in\{1,2,3\}$ denotes the spatial dimension. As shown in FIG. \ref{fig11}, the objective of our approach is to explore Green's function of the operator, $\mathcal{L}$, using $N$ input-output pairs $\left\{\left(f_j, u_j\right)\right\}_{j=1}^N$, consisting of forcing functions, $f_j: \Omega \rightarrow \mathbb{R}$, and system responses, $u_j: \Omega \rightarrow \mathbb{R}$, which satisfy the following equation:
\begin{equation}
\left\{
\begin{aligned}
    \mathcal{L} u_j &= f_j, \\
    \mathcal{D}\left(u_j, \Omega\right) &= g.
\end{aligned}
\right.
\label{eq1}
\end{equation}
where $\mathcal{D}$ is a linear operator acting on the solutions $u$, and the domain $\Omega$; with $g$ represents the constraint. We assume that the forcing terms possess adequate regularity, and that the operator $\mathcal{D}$ imposes a constraint ensuring Eq. (\ref{eq1}) has a unique solution. For example, such a constraint could be the imposition of homogeneous Dirichlet boundary conditions on the solutions: $\mathcal{D}\left(u_j, \Omega\right):=\left.u_j\right|_{\partial \Omega}=0$. It is worth noting that various constraints such as boundary conditions, integral conditions, jump conditions, or non-standard constraints are all plausible.

A Green's function of the operator $\mathcal{L}$ is defined as the solution to the following equation:
\begin{equation}
\left\{
\begin{aligned}
    \mathcal{L} G(x, y) &=\delta(y-x), &\quad x \in \Omega, \\
    \mathcal{D}G(x, y) &= g, &\quad x \in \partial\Omega.
\end{aligned}
\right.
\label{eq2}
\end{equation}
where $\mathcal{L}$ acts on the function $x \mapsto G(x, y)$ for fixed $y \in \Omega$, and $\delta(\cdot)$ represents the Dirac delta function. The Green's function is well-defined and unique under mild conditions on $\mathcal{L}$, and suitable solution constraints imposed by an operator, $\mathcal{D}$. Moreover, if $(f, u)$ is an input-output pair satisfying Eq. (\ref{eq1}) with $g=0$, then:
\begin{equation}
    u(x)=\int_{\Omega} G(x, y) f(y) d y, \quad x \in \Omega .
\end{equation}

Therefore, the Green's function associated with $\mathcal{L}$ can be thought of as the right inverse of $\mathcal{L}$.
Let $u_{\text{hom}}$ be the homogeneous solution to Eq. (\ref{eq1}), so that:
\begin{equation}
\left\{
\begin{aligned}
    \mathcal{L} u_{\text{hom}}&=0, \\
    \mathcal{D}\left(u_{\text{hom}}, \Omega\right)&=g.
\end{aligned}
\right.
\end{equation}

Using superposition, we can construct solutions, $u_j$, to Eq. (\ref{eq1}) as $u_j=\tilde{u}_j+u_{\text{hom}}$, where $\tilde{u}_j$ satisfies:
\begin{equation}
\left\{
\begin{aligned}
    \mathcal{L} \tilde{u}_j&=f_j, \\
    \mathcal{D}\left(\tilde{u}_j, \Omega\right)&=0.
\end{aligned}
\right.
\label{eq5}
\end{equation}

Then, the relation between the system's response, $u_j$, and the forcing term, $f_j$, is expressed via the Green's function as:
\begin{equation}
u_j(x)=\int_{\Omega} G(x, y) f_j(y) d y+u_{\text{hom}}(x), \quad x \in \Omega.
\label{eq6}
\end{equation}

Therefore, we use DISCOVER, to find the Green's function ($\mathcal{T}: \Omega \times \Omega \rightarrow \mathbb{R} \cup\{ \pm \infty\}$) associated with $\mathcal{L}$ and the constraint operator $\mathcal{D}$ in Eq. (\ref{eq5}). The schematic of learning Green's function based on DISCOVER is illustrated in FIG. \ref{fig11}. The discovered Green's functions can provide insights into the physical properties of the operator $\mathcal{L}$ and the types of boundary constraints imposed.

\paragraph*{\textbf{Generation of training data.}}
The training dataset consists of $N$ forcing functions, $f_j: \Omega \rightarrow \mathbb{R}$, and associated system responses, $u_j: \Omega \rightarrow \mathbb{R}$, which are solutions to Eq. (\ref{eq1}):
Unless otherwise stated, the training data comprises $N=100$ pairs of forcing and solution, where the forcing terms are randomly drawn from a Gaussian process $\mathcal{G} \mathcal{P}\left(0, K_{\mathrm{SE}}\right)$, where $K_{\mathrm{SE}}$ is the squared-exponential covariance kernel \cite{williams2006gaussian} defined as
\begin{equation}
    K_{\mathrm{SE}}(x, y)=\exp \left(-\frac{|x-y|^2}{2 \ell^2}\right), \quad x, y \in \Omega .
    \label{eq7}
\end{equation}
The parameter $\ell>0$ in Eq. (\ref{eq7}) is called the length-scale parameter, and characterizes the correlation between the values of $f \sim \mathcal{G}\mathcal{P}\left(0, K_{\mathrm{SE}}\right)$ at $x$ and $y$ for $x, y \in \Omega$. A small parameter, $\ell$, yields highly oscillatory random functions, $f$, and determines the ability of the Gaussian process to generate a diverse set of training functions. This property is crucial for accurately capturing different modes within the operator, $\mathcal{L}$, and learning the associated Green's function \cite{boulle2023learning}. 
When Eq. (\ref{eq1}) is a boundary-value problem, we generate 100 training pairs by solving the PDEs with a spectral method 
 \cite{trefethen2000spectral} using the Chebfun software system 
 \cite{driscoll2014chebfun}, written in MATLAB, and using a tolerance of $5 \times 10^{-13}$.

\subsection{Symbolic regression with physical hard constraints based on DISCOVER.}

\paragraph*{\textbf{Symbolic mathematical representation of open-form expressions.}}
In this paper, our aim is to discover the Green's function with the following form:
\begin{equation}
    \hat{G}=\mathcal{T}(x,y; \xi), \quad \mathcal{T}: \mathbb{R}^D \rightarrow \mathbb{R}
    \label{eq8}
\end{equation}
where the state variables $x, y \in \mathbb{R}^D$, and $\xi$ denotes possible constants. $\hat{G}$ refers to the possible Green's function. Our goal is to find an optimal expression $\mathcal{T}$ that accurately describes the true underlying physical laws in Eq. (\ref{eq6}), while maintaining the concise form of $\hat{G}$.
In order to realize automatic mining for unknown Green's function, the key
issue is how to represent diverse open-form expressions in a flexible manner.
From the perspective of symbolic mathematics, any equation can be expressed as a graph \cite{atkinson2019data} or a binary tree \cite{lample2019deep}, which might be a feasible representation of open-form candidate set. 

In this paper, symbolic binary trees are taken as a representation of function terms and each open-form expression is regarded as a forest. 
Specifically, operations involving two objects are defined as binary operators (e.g., addition, subtraction, multiplication, and division). The operations involving a single object, such as exponents, logarithms, and trigonometric functions, are defined as unary operators. Operands are variables in the equation, both independent and dependent, such as $x$ and $y$. In binary tree structure, all leaf nodes represent operands since they could have two degree/children, while internal nodes correspond to operators since they only have one degree/children. 

Moreover, each expression tree has a unique preorder traversal sequence corresponding to it. Therefore, we can conveniently generate batches of preorder traversal sequences by means of the LSTM agent of DISCOVER, instead of the expression trees.

\begin{figure}[H]
    \centering
    \includegraphics[width=0.9\textwidth]{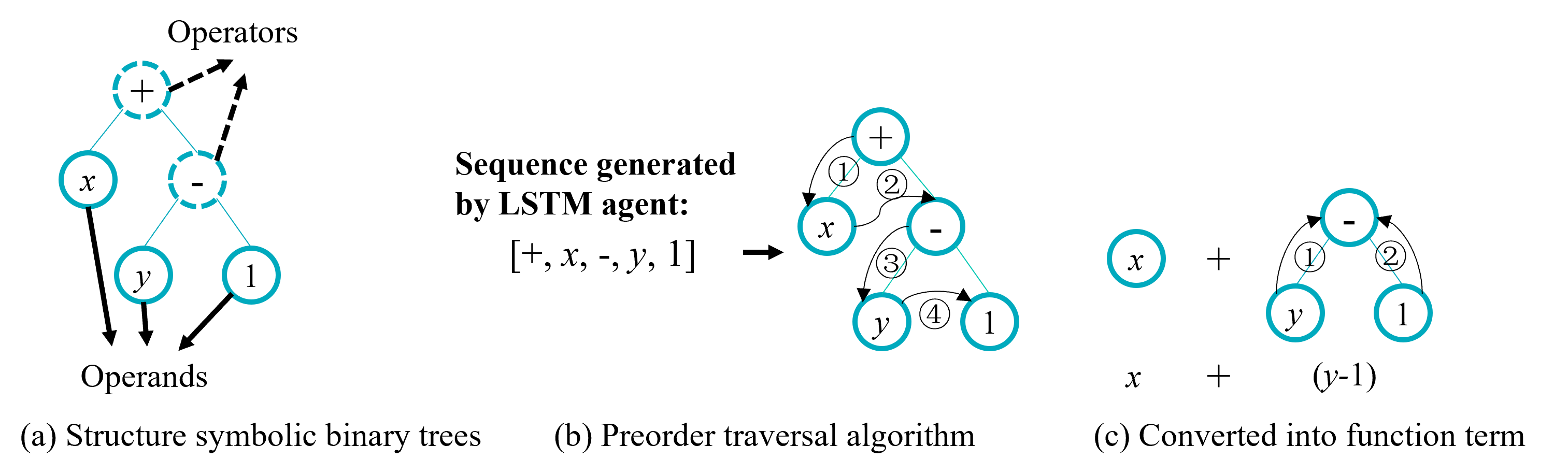}
        \caption{Schematic diagram of binary tree representation and transformation method based on preorder traversal algorithm.}
    \label{fig12}
\end{figure}

The transformation from the preorder traversal sequence generated by LSTM agenet in DISCOVER to the binary tree is based on preorder traversal algorithm. Speciffically, as shown in FIG. \ref{fig12} (b), preorder traversal begins at the root node of the binary tree. During computation, the left child node is processed first, and the traversal continues layer by layer to the leaf nodes. Next, we backtrack to the nearest parent node with an incomplete degree, proceed to the other branch (guided by the right child node), and continue traversing to its leaf node. The above process is repeated until all internal nodes of the entire binary tree are traversed. FIG. \ref{fig12} (b) shows a binary tree corresponding to the function term $x+(y-1)$, and the black lines show the path of the preorder traversal. For more examples, FIG. \ref{fig1} (c) displays a binary tree corresponding to the function term $x(y-const)$. 

As FIG. \ref{fig12} (c) demonstrated, the binary tree is then converted into mathematical expression based on subtree division and postorder traversal manner. Firstly, the binary tree is spilt into subtrees (i.e., function terms) based on the plus and minus operators at the top of the expression tree. 
Subsequently, the value of each function term is solve based on postorder
traversal manner. Taken $(y-1)$ as an example in FIG. \ref{fig12} (c), the subtree is traversed from bottom to top ($\textcircled{1} \rightarrow \textcircled{2}$) and then perform operations at each parent node (operators). Therefore, the mathematical expression can be acquired and the value can be calculated during this process.

\paragraph*{\textbf{DISCOVER framework.}}
The DISCOVER framework \cite{du2024discover} is utilized in this paper. 
Compared with other symbolic regression methods, DISCOVER utilizes a structure-aware agent to efficiently produce mathematical expressions with a tree structure. 
The symbol library is easily expandable.
Customized constraints and regulations can be included to expedite the search process of deep reinforcement learning methods based on prior knowledge.

Unveiling the Green's function via the DISCOVER framework involves three key phases: (1) generate expressions with a tree structure based on a pre-defined symbol library by LSTM agent; (2) determine the value of the constants; and (3) evaluate acquired expressions by a user-designed reward function. The whole process is iteratively update by the deep reinforcement learning methods until best-case expression meets the accuracy and parsimony requirements that we set in advance. 

Details of each step are elaborated as follows:

\noindent (1) In the generation phase, we initiate by constructing a pre-defined symbol library encompassing elementary functions, including $+$, -, $\times$, $\div$, $\sqrt{\cdot}$, $e$, $\ln$, sin, cos, arcsin, and tanh, along with corresponding rules. 
For instance, the prohibition of more than ten instances of $+$. Constants could not be the sole child of fundamental function operators. sin/cos/$e$/$\ln$ could not be the descendants of sin/cos/$e$/$\ln$. 
The LSTM agent then iteratively generates diverse potential expressions while adhering to aforementioned predefined constraints through dirtribution probability multiplication (details can be found in paper\cite{du2024discover}) [as in FIG. \ref{fig1} (b)]. 

\noindent (2) Except for the operators, the constants are also crucial in the formulation of the Green's function. Specially, in this paper, considering the complexity and uncertainty of the Green's function expressions, the constants $\xi$ can appear at any position in the expression tree. To tackle this problem, we first generate equation skeleton through LSTM agents of DISCOVER with constants taken as placeholders, and then utilize the BroydenFletcher-Goldfarb-Shanno algorithm (BFGS) \cite{HEAD1985264} to execute the following optimization objective. By this way, the value of the constant terms and the specific expressions can be automatically determined through Eq. (\ref{eq9})-(\ref{eq10}). 

\begin{equation}
\xi^*=\operatorname{argmin}_{\xi} \sum_{j=1}^n \frac{1}{n}\left(\hat{u}_j-u_j\right)^2,
\label{eq9}
\end{equation}

\begin{equation}
    \hat{u}_j = 
    \int_{\Omega} \mathcal{T}(x,y; \xi) f_j(y) d y+u_{\text{hom}}(x), \quad x \in \Omega.
\label{eq10}
\end{equation}

Note that, in Eq. (\ref{eq9})-(\ref{eq10}), $\hat{u}$ is calculated by the integral with learned Green's function $\mathcal{T}(x,y; \xi)$ and source term $f$, while the exact $u$, which is training data, is pre-calculated by solving Eq. (\ref{eq1}) using a spectral method \cite{trefethen2000spectral}. Rounds of optimization iterations are performed to ultimately determine all the constants $\xi$ in the expression tree $\mathcal{T}(x,y; \xi)$.

\noindent (3) During the evaluation and optimization phase, a deep reinforcement learning approach is employed to maximize the reward function. The reward function is user-defined and details of the definition of reward function will be given in latter.
It should be note that, the search space comprises a discrete set of various mathematical expression combinations, and thus the nondifferentiable relationship between the rewards and the agent’s parameters renders traditional optimization methods based on gradient descent inapplicable. That's why the deep reinforcement-learning training strategy is adopted in DISCOVER. 

\begin{figure}[H]
    \centering
    \includegraphics[width=0.7\textwidth]{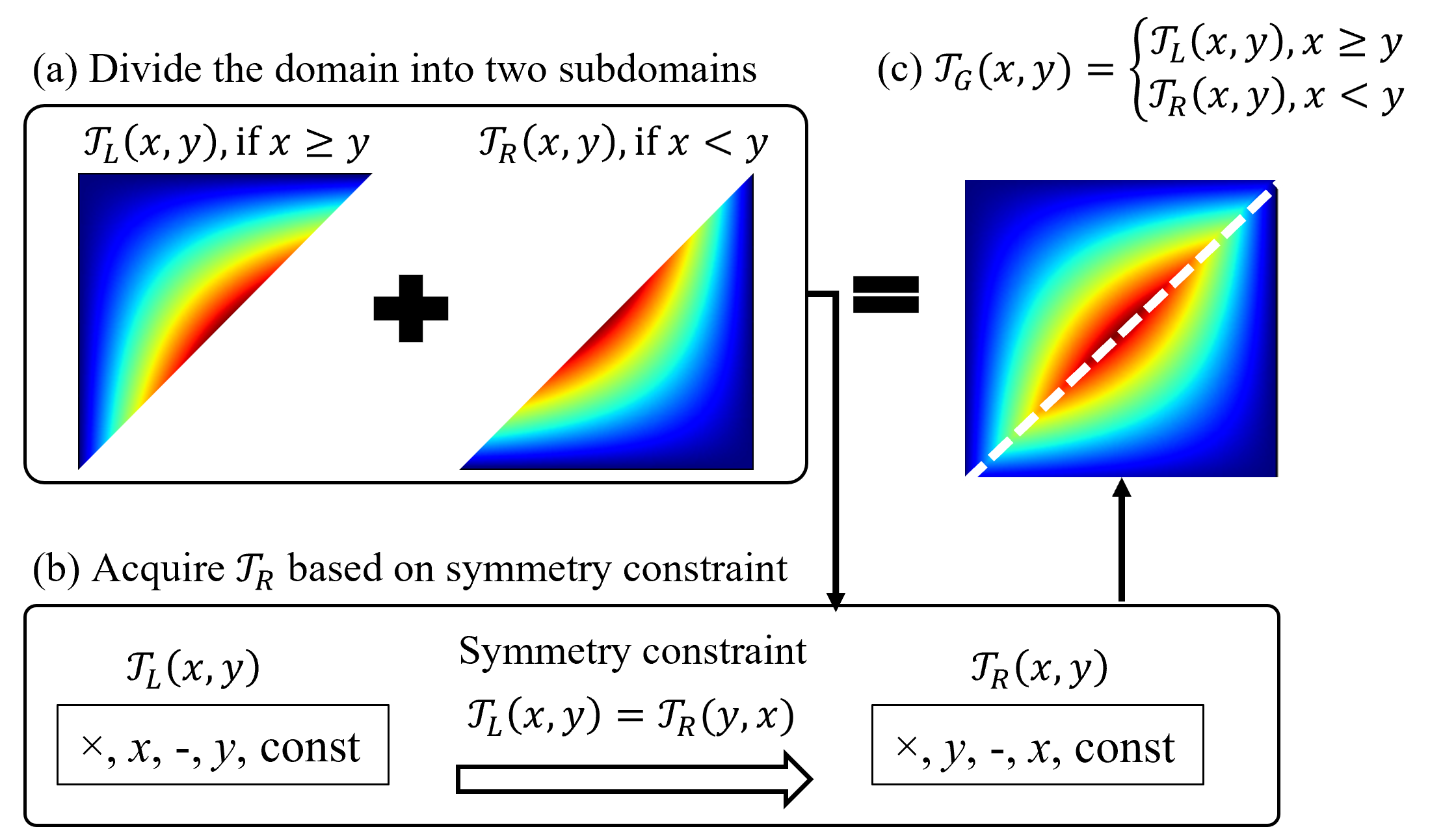}
        \caption{Schematic diagram of symmetry hard constraint.}
    \label{fig13}
\end{figure}

\paragraph*{\textbf{Physical hard constraints.}}
If the differential operator is self-adjoint, the associated Green's function exhibits symmetry, expressed as $G(x, y) = G(y, x)$ for all $x, y \in \Omega$, contingent upon the constraint operator $\mathcal{D}$. As FIG. \ref{fig13} (a) shown, the symmetry can be represented as a combination of two subdomains divided by line $y=x$ (the whit dotted line), with the constraint of the two subdomains satisfied symmetrical property. To facilitate this symmetry, we thuse devise a symmetric tree denoted as $\mathcal{T}(x, y)$, comprising a left subtree $\mathcal{T}_L(x, y)$ and a right subtree $\mathcal{T}_R(x, y)$. 
The relationship among $\mathcal{T}(x, y)$, $\mathcal{T}_L(x, y)$, and $\mathcal{T}_R(x, y)$ is characterized by the following equations:

\begin{equation}
\left\{
\begin{aligned}
    \mathcal{T}(x, y) &= \mathcal{T}_L(x, y)\cdot(x\geq y)+\mathcal{T}_R(x, y)\cdot(x<y), \\
    \mathcal{T}_L(x, y) &= \mathcal{T}_R(y, x).
\end{aligned}
\right.
\end{equation}

Specifically, we conducted this symmetry constraint in the generation phase of DISCOVER. After the tree is produced by the LSTM agents, we define it as a left subtree $\mathcal{T}_L(x, y)$. As FIG. \ref{fig1} (d) and FIG \ref{fig13} (b) shown, the right subtree $\mathcal{T}_R(x, y)$ can be constituted by simultaneously replacing $x$ with $y$ and replacing $y$ with $x$ in the traversal of the left subtree. In this way, the constraint of $\mathcal{T}_L(x, y) = \mathcal{T}_R(y, x)$ can be naturally satisfied. 
Take the Green's function of the the Laplace operator as an example, if a preorder traversal $[\times,x,+,y,const]$ is produced by the LSTM agent, the left subtree $\mathcal{T}_L(x, y)$ can be represented as $x(y+const)$. Therefore, the traversal of the right subtree can be obtained as $[\times,y,+,x,const]$ through the substitution according to the symmetry constraint. The expression $\mathcal{T}_R(x, y)$ can then be ascertained as $y(x+const)$. Finally, the expression of $\mathcal{T}(x, y)$ is constructed as $x(y+const)\cdot(x\geq y)+y(x+const)\cdot(x<y)$, where the symmetry of $\mathcal{T}(x, y)= \mathcal{T}(y, x)$ can be satisfied for all $x, y \in \Omega$.

In this sense, the symmetry property of the Green's function can be achieved through the imposition of hard constraints. Compared with soft constraints, such as penalty terms incorporated within loss functions, hard constraints ensure rigorous adherence to symmetry, thereby bolstering predictability, stability, and reliability. We conduct an ablation study in section \ref{sec3} to elucidate the efficacy of symmetric hard constraints. Specifically, to assess the influence of this term on the learning outcome, we employ the DISCOVER framework in two distinct manners: one integrates symmetric hard constraints during the generation phase of symbolic trees, and the other omits such constraints altogether.

\paragraph*{\textbf{Reward function.}}
To effectively assess the suitability of generated expression candidates, we devise a reward function for the Green's discovery problem that holistically considers observation fitness and equation simplicity. The formulation is given as:
\begin{equation}
R = \frac{1 - \zeta_1 \times n - \zeta_2 \times D}{1 + \text{MSE}},
\label{eq12}
\end{equation}
where $n$ denotes the number of function terms in the governing equation, $D$ represents the depth of the generated expression tree, and $\zeta_1$ and $\zeta_2$ are penalty factors for equation simplicity and are typically set to small, untuned values. In our default hyperparameter configuration, these penalty factors are consistently assigned as 0.01 and 0.0001, respectively, and remain unchanged across all numerical experiments. Here, the mean square error (MSE) is utilized to quantify the discrepancy between predicted and observed values. The term $\frac{1}{(1 + \text{MSE})}$ normalizes the reward value within a range of 0 to 1, facilitating the interpretation of the model performance. A reward value of 1 indicates precise equivalence between the equation's sides, thereby minimizing error. This design choice mitigates challenges associated with sparse reward distribution, a common issue in symbolic regression tasks \cite{sun2022symbolic, petersen2019deep}.

Inspired by the PINN framework \cite{raissi2020hidden}, the MSE is computed as:
\begin{equation}
    \text{MSE} =\frac{1}{N} \sum_{j=1}^N \int_{\Omega}\left(u_j(x)-u_{\text {hom}}(x)-\int_{\Omega} \mathcal{T}(x, y; \xi) f_j(y) \mathrm{d} y\right)^2 \mathrm{~d} x.
    \label{eq13}
\end{equation}
where $N$ is the number of observation pairs, $u_j(x)$ represents the observed solution, $u_{\text{hom}}(x)$ denotes the homogeneous solution, and $\mathcal{T}(x, y)$ is the Green's function.

\section{Experiments and results.\label{sec3}}
In this section, to validate the effectiveness of our method, we conducted a two-part experiment: (i) discover operators that have ground-truth Green’s functions (i.e. Laplace operator and Helmholtz operator); (ii) discover operators without ground-truth Green’s functions (i.e. the periodic Helmholtz operator and the second-order differential operator with jump conditions). The first part experiment is conducted to show the accuracy of our method while the second part experiment is carried out to demonstrate that this method can genuinely discover new knowledge in unknown scenarios.

Specifically, for the (i) experiments involving operators with ground-truth Green’s functions, we compared the discovered Green’s functions with the ground-truth versions, as well as the differences in PDE solutions, to verify the accuracy of our method. We also compared our method with similar approaches (PySR and rational neural network proposed by Boulle et al. \cite{boulle2022data}) to demonstrate its superiority. It is shown that the discovered operator for Laplace operator and Helmholtz operator is highly consistent with the ground-truth Green's functions, which surpass the performance of PySR and neural network. 

For the (ii) experiments involving operators lacking ground-truth Green’s functions, we validated the accuracy of our method by examining the differences in PDE solutions and assessed the reasonableness of the discovered Green’s functions based on prior physical knowledge. The results demonstrated a minute solution errors on the order of $10^{-10}$, which is hard to achieve by approximation methods, including neural network.

Finally, we conducted noise experiments to verify the robustness of our method. Moreover, ablation studies methodically eliminate or deactivate system components to assess their specific impacts and overall significance on system performance. We performed ablation studies in this paper to showcases the effectiveness of the applied symmetry constraints.

The unearthed Laplace and Helmholtz operators are crucial for addressing the pressure Poisson equation, and the identified second-order differential operators with discontinuous conditions are indispensable for examining multiphase flows and shock dynamics. The application of symbolic regression to unearth Green's functions marks a significant stride in harnessing AI to expedite scientific insights, especially within fluid mechanics and cognate disciplines.

The default hyperparameters employed in mining latent Green's functions are detailed in Table \ref{table1}. The parsimony penalty factor, crucial for ensuring equation simplicity, is set to 0.01. Throughout the optimization process, the coefficients for entropy loss and policy gradient loss are fixed at 0.03 and 1, respectively. The maximum depth of the generated subtree is set as 10. During each iteration, the agent generates a total of $N_{\text{expression}} = 500$ expressions. Maximum iteration number is set as 100.

\begin{table}[H]
    \centering
    \caption{Default hyperparameter settings for DISCOVER framework.}
    \begin{tabular}{ccc}
    \toprule
    Hyperparameter & Default value & Definition \\
    \hline
    $D_{\text{subtree}}$ & 10 & Maximum depth of subtrees \\
    $\zeta_1$ & 0.01 & Parsimony penalty factor for redundant function terms \\
    $\zeta_2$ & 0.0001 & Parsimony penalty factor for unnecessary structures \\
    $N_{\text{expression}}$ & 500 & Total generated expressions at each iteration \\
    $N_{\text{epoch}}$ & 100 & Total iteration number\\
    \bottomrule
    \end{tabular}
    \label{table1}
\end{table}

\subsection{Discovery results compared with ground-truth Green's functions.}
\subsubsection*{Case 1: Laplace operator.}
In fluid dynamics, the Laplacian operator is critically important for describing the diffusion of quantities such as velocity, temperature, and concentration within a fluid. It appears prominently in the Navier-Stokes equations, which govern the motion of viscous fluid substances. The Laplacian helps to model the rate at which these quantities spread out over time, contributing to our understanding of phenomena like heat conduction, viscous dissipation, and the smoothing of velocity fields. Its application is essential for accurately simulating and analyzing fluid behavior in various engineering and scientific contexts.
In this case, we attempt to learn the Green's function of the Laplace operator on $\Omega = [0,1]$, with homogeneous Dirichlet boundary conditions defined as Eq. (\ref{eq14}). The ground-truth Green's function for $x, y \in[0,1]$, is given by Eq. (\ref{eq15}). As stated in section \ref{sec2}, we used 100 input-output pairs $\left\{\left(f_j, u_j\right)\right\}_{j=1}^N$ acquired by Gaussian process with squared-exponential covariance kernel. Note that the exact Green's function is not fed into the discovering process.

\begin{equation}
\left\{
\begin{aligned}
    \mathcal{L} u &=-\frac{d^2 u}{d x^2}=f, \quad &x \in \Omega,\\
    \mathcal{D} u &=0, \quad &x \in \partial\Omega.
\end{aligned}
\right.
\label{eq14}
\end{equation}

\begin{equation}
    G_{\text{exact}}(x, y)= \begin{cases}x(1-y), & \text { if } x \leq y, \\ y(1-x), & \text { if } x>y.\end{cases}
    \label{eq15}
\end{equation}

\begin{figure}[H]
    \centering
    \includegraphics[width=1\textwidth]{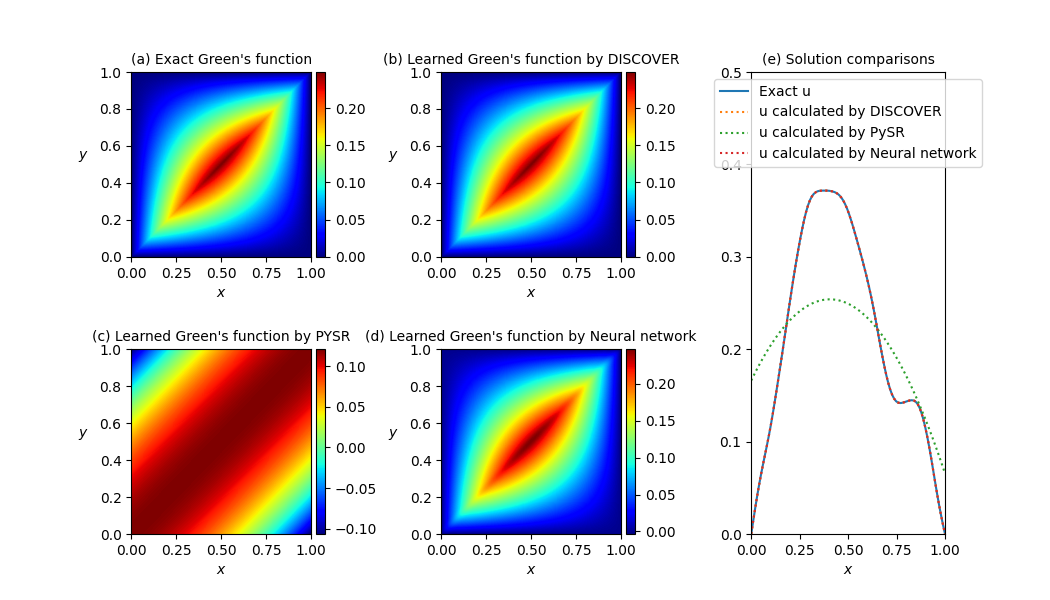}
    \caption{Laplace operator. (a) Exact Green's function. (b) Learned Green’s functions by DISCOVER and (c) PySR and (d) rational neural network (black box models). (e) Solution comparisons with exact $u$, $u$ calculated by DISCOVER, $u$ calculated by PySR, and $u$ calculated by rational neural network (black box models).}
    \label{fig2}
\end{figure}

 The Green's function found by DISCOVER is:
\begin{equation}
\mathcal{T}(x, y)=\left\{
\begin{aligned}
    \mathcal{T}_L(x, y) &= 0.999995x-0.999986xy, & \text { if } x \leq y,\\
    \mathcal{T}_R(x, y) &= 0.999995y-0.999986xy. & \text { if } x>y.
\end{aligned}
\right.
\end{equation}
The reward of the learned Green's function calculated by Eq. (\ref{eq9}) is 0.9899, whilst the MSE error calculated by Eq. (\ref{eq10}) is $1.4797\times 10^{-10}$. The learned Green's function closely approximates the exact Green's function, achieving qualification within just 2 epochs using the DISCOVER framework, underscoring the efficiency of our approach. The calculated solution $\hat{u}$ by learned Green's function of DISCOVER is demonstrated in FIG. \ref{fig2} (e), which is exactly closed to the $u$ calculated by spectral method.

In comparison, results obtained via PySR \cite{cranmer2023pysr} and rational neural network method proposed by Boulle et al. \cite{boulle2022data} are also presented. PySR \cite{cranmer2023pysr} is an open-source library for practical symbolic regression. By leveraging a powerful multi-population evolutionary algorithm, a unique evolve-simplify-optimize loop, a high-performance distributed backend, PySR is capable of accelerating the discovery of interpretable symbolic models from data. 
The training data set of PySR, the maximum depth of subtrees, the total generated expressions at each iteration, and total iteration number are keep same as that of DISCOVER to ensure fairness. The expression yielded by PySR is provided in Eq. (\ref{eq17}), exhibiting a MSE error of $7.4230\times 10^{-3}$. Albeit the symmetry is also satisfied in Eq. (\ref{eq17}), the expression is greatly different from the exact expression in Eq. (\ref{eq15}). The calculated solution $\hat{u}$ by learned Green's function of PySR (ref FIG. \ref{fig2} (e)) is also evidently different from the actual $u$.
Notably, the DISCOVER framework demonstrates superior performance in extracting the Green's function, consequently ensuring an exact solution $\hat{u}$. 

\begin{equation}
\mathcal{T}_{\text{PySR}}(x, y) = -0.056956 \times (2x-2y)^2 + 0.12115.
\label{eq17}
\end{equation}

Unlike symbolic regression based methods, rational neural network is a kind of black box method with powerful approximation ability.  
Rational neural network consists of neural network with adaptive rational activation functions $x \mapsto \sigma(x)=p(x) / q(x)$, where $p$ and $q$ are two polynomials, whose coefficients are trained at the same time as the other parameters of the networks, such as the weights and biases. These coefficients are shared between all the neurons in a given layer but generally differ between the network's layers. This type of network was proven to have better approximation power than standard Rectified Linear Unit (ReLU) networks, which means that they can approximate smooth functions more accurately with fewer layers and network parameters. The acquired Green's function by rational neural network is shown in FIG. \ref{fig2} (d), which is exactly same as exact Green's function (FIG. \ref{fig2} (a)) and Green's function learned by DISCOVER (FIG. \ref{fig2} (c)). However, the number of weight parameters in the Green's function found by DISCOVER is just 2, whereas that by the rational neural network is 5253. Compared to DISCOVER, the rational neural network requires a significantly larger number of parameters to approximate the Green's function, which poses considerable inconvenience for further understanding, analyzing, and applying the obtained Green's function. Moreover, considering that neural networks are essentially black-box models, the correctness of the Green's function obtained through neural network training remains to be further examined.

\begin{figure}[H]
    \centering
    \includegraphics[width=0.85\textwidth]{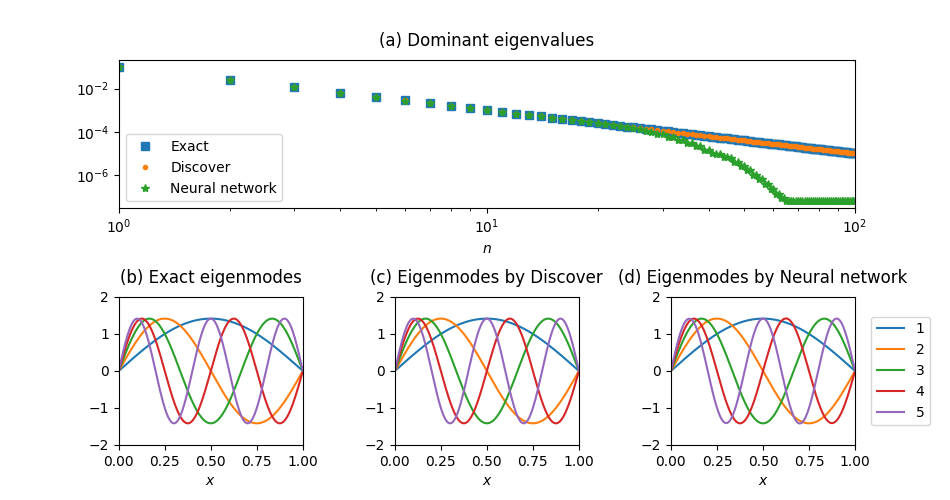}
    \caption{Eigenvalue decomposition of Laplace operator. (a) The first 100 largest eigenvalues of the exact Green's function, the learned Green’s functions by DISCOVER and learned Green’s functions by neural network (black box models). The first five eigenfunctions of (b) the exact Green’s functions, (c) learned Green’s functions by DISCOVER, (d) learned Green’s functions by neural network (black box models).}
    \label{fig7}
\end{figure}

For further validation, we turn to the eigenvalue decomposition of the Green's function. The eigenvalues of PDE operators are crucial for determining the stability, behavior, and properties of solutions. They play a fundamental role in various contexts such as stability analysis, modal decomposition, quantization in quantum mechanics, and the analysis of heat, diffusion, and wave propagation phenomena. As delineated in literature \cite{stakgold2011green}, the differential operator $\mathcal{L}$ and the integral operator with kernel $G$ share the same eigenfunctions but possess reciprocal eigenvalues.  Here, we assess our method's capability to accurately recover the eigenfunctions of the Green's function associated with the largest eigenvalues in magnitude, from input-output pairs. As illustrated in FIG. \ref{fig7} (a), despite slight differences in the coefficients of the discovered Green's function by DISCOVER compared to the ground-truth Greens function, both the eigenvalues and dominant eigenmodes are well recovered. Notably, while neural networks demonstrate performance comparable to that of the DISCOVER framework as depicted in FIG. \ref{fig2}, they exhibit an decay of the eigenvalues of the learned Green's function. This observation may be attributed to the neural network's smooth approximation of the exact Green's function \cite{boulle2022data}. The eigenvalues of the Green's function approximated by the rational neural network match the actual eigenvalues only for the first 30 values, whereas the eigenvalues of the Green's function discovered by DISCOVER match the actual eigenvalues completely. This has significant implications for subsequent stability analysis and modal decomposition of the equation. In this sense, the DISCOVER framework is deemed better suited for scientific discovery, as it can provide more fundamental physical core information.

\subsubsection*{Case 2: Helmholtz operator with Dirichlet boundary conditions.}
In fluid dynamics, the Helmholtz operator is fundamental for analyzing and solving problems related to wave propagation, acoustics, and fluid flow. It appears in the Helmholtz equation, which describes how physical fields like pressure, velocity potential, and acoustic pressure behave under steady-state conditions. The Helmholtz operator's role is crucial in decomposing complex flow fields into vortices and potential flows, facilitating the study of turbulent flows, aeroacoustics, and the stability of fluid systems. Its application is essential for understanding phenomena such as sound wave propagation in fluids and the behavior of incompressible flows.
In this example, we aim to learn the Green’s function of a high frequency Helmholtz operator with homogeneous Dirichlet boundary conditions on $\Omega=[0,1]$:
\begin{equation}
\left\{
\begin{aligned}
    \mathcal{L} u &=\frac{d^2 u}{d x^2}-K^2 u=f, \quad &x \in \Omega,\\
    \mathcal{D} u &=0, \quad &x \in \partial\Omega.
\end{aligned}
\right.
\end{equation}
where $K=8$ denotes the Helmholtz frequency. As stated in section \ref{sec2}, we used 100 input-output pairs $\left\{\left(f_j, u_j\right)\right\}_{j=1}^N$ acquired by Gaussian process with squared-exponential covariance kernel. 

The analytic expression, $G_{\text {exact}}$, is given by
\begin{equation}
    G_{\text {exact}}(x, y)= \begin{cases}\frac{\sinh (K x) \sinh (K(y-1))}{K \sinh (K)}, & \text { if } x \leq y, \\ \frac{\sinh (K y) \sinh (K(x-1))}{K \sinh (K)}, & \text { if } x>y,\end{cases}
    \label{eq19}
\end{equation}
where $x, y \in[0,1]$. Note that the exact Green's function is not fed into the discovering process.

The expression extracted by DISCOVER is:
\begin{equation}
\mathcal{T}(x, y)=\left\{
\begin{aligned}
    \mathcal{T}_L(x, y) &= 8.272347\times10^{-5}\cdot sinh\left(-8.015867 x\right) \cdot sinh(8.014433-8.015321y), & \text { if } x \leq y,\\
    \mathcal{T}_R(x, y) &= 8.272347\times10^{-5}\cdot sinh\left(-8.015867 y\right) \cdot sinh(8.014433-8.015321x). & \text { if } x>y,
\end{aligned}
\right.
\end{equation}

\begin{figure}[H]
    \centering
    \includegraphics[width=0.95\textwidth]{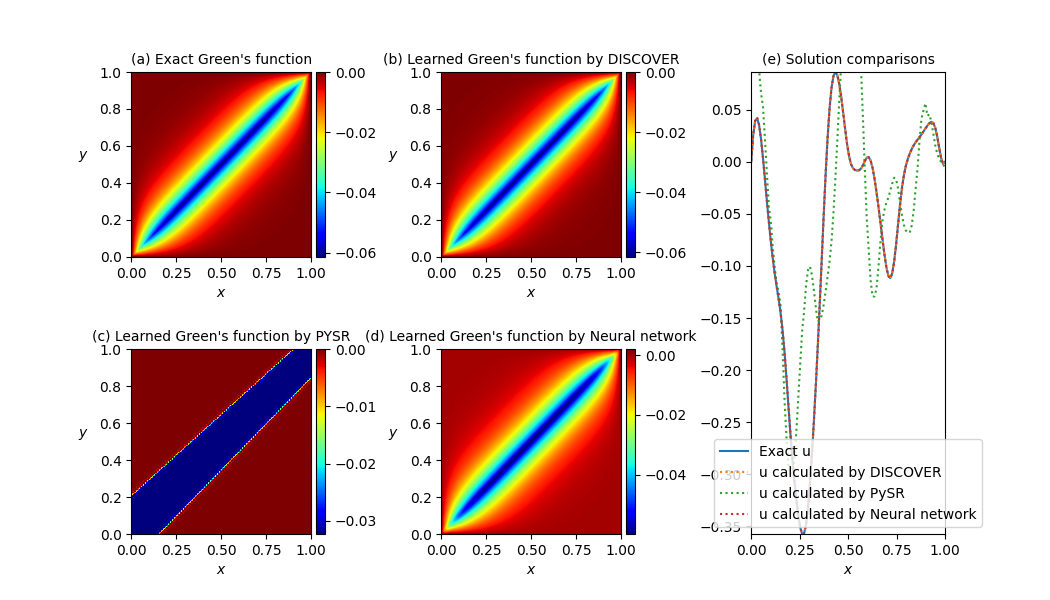}
    \caption{Helmholtz operator. (a) Exact Green's function. (b) Learned Green’s functions by DISCOVER and (c) PySR and (d) rational neural network (black box models). (e) Solution comparisons with exact $u$, $u$ calculated by DISCOVER, $u$ calculated by PySR, and $u$ calculated by rational neural network (black box models).}
    \label{fig3}
\end{figure}

The reward calculated for the learned Green's function, as per Eq. (\ref{eq9}), stands at 0.9891, while the MSE computed via Eq. (\ref{eq10}) amounts to $5.0990\times 10^{-8}$. Interestingly, in Eq. (\ref{eq19}), the coefficient $\frac{1}{K \sinh (K)}=\frac{1}{8 \sinh (8)} \approx 8.3866\times10^{-5}$, which is very close to the coefficient of the learned Green's function by DISCOVER. It is worth noting that the coefficients of the Green's function for this high-frequency operator are extremely small (on the order of $10^{-5}$). However, DISCOVER is still able to accurately learn the implicit Green's function embedded within the data. This underscores the remarkably strong knowledge discovery capability inherent in the DISCOVER framework.
FIG. \ref{fig3} presents the learned Green's function alongside its associated homogeneous solution. Notably, the expressions of $\mathcal{T}(x, y)$ and $G_{\text{exact}}(x, y)$, as well as their respective images and computed numerical solutions, exhibit high consistency.
Consistent with the expectations outlined in Eq. (\ref{eq2}), it is observed that $\mathcal{T}(0, y)=0$ and $\mathcal{T}(1, y)=0$ for all $y \in[0,1]$. Additionally, in accordance with the symmetry property of the Green's function, it holds true that $\mathcal{T}(x, 0)=0$ and $\mathcal{T}(x, 1)=0$ for all $x \in[0,1]$.

In comparison, similar to \textbf{Case 1}, results obtained via PySR \cite{cranmer2023pysr} and rational neural network method proposed by Boulle et al. \cite{boulle2022data} are also presented. To maintain fairness, the PySR training dataset, maximum subtree depth, total expressions generated per iteration, and total iteration number were aligned with those used in DISCOVER. The expression produced by PySR, shown in Eq. (\ref{eq21}), has a MSE of $3.356\times 10^{-2}$. This expression significantly deviates from the exact form given in Eq. (\ref{eq19}). The solution $\hat{u}$ derived from the Green's function learned by PySR (see FIG. \ref{fig3} (e)) also shows substantial discrepancies compared to the actual $u$. Importantly, the DISCOVER framework exhibits superior performance in extracting the Green's function, thereby ensuring an accurate solution $\hat{u}$.

\begin{equation}
\mathcal{T}_{\text{PySR}}(x, y) = -0.02207 \sinh\left[\sinh((0.392895 + y)\geq (x + 0.2370032))\right]\cdot \left[x \geq (1.1423628y - 0.2370032)\right].
\label{eq21}
\end{equation}

The acquired Green's function by rational neural network is shown in FIG. \ref{fig3} (d), which is exactly same as exact Green's function (FIG. \ref{fig3} (a)) and Green's function learned by DISCOVER (FIG. \ref{fig3} (c)). The solution $\hat{u}$ is also almost identical to exact solution $u$. However, as stated before, the main drawback of the rational neural network is the lack of explicit expression for the Green's function, and thus the correctness of the Green’s function obtained through neural network training remains to be further examined.

\begin{figure}[H]
    \centering
    \includegraphics[width=0.85\textwidth]{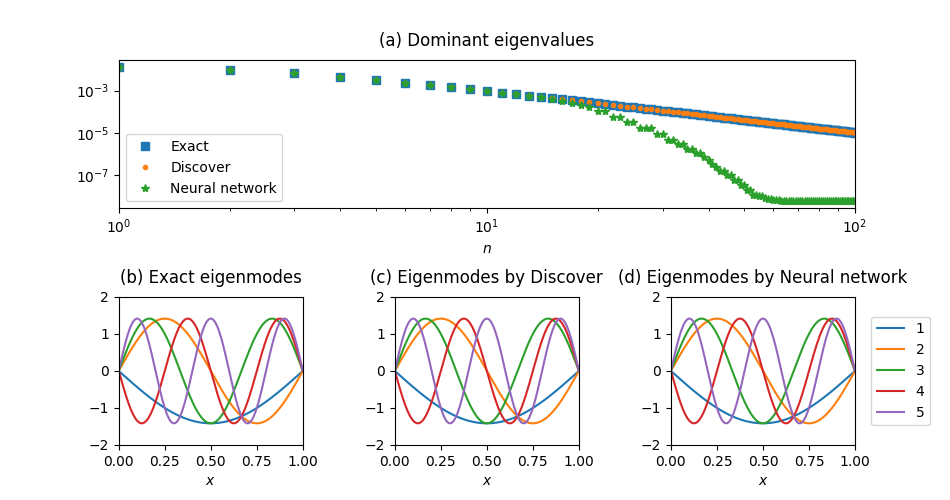}
    \caption{Eigenvalue decomposition of Helmholtz operator. (a) The first 100 largest eigenvalues of the exact Green's function, the learned Green’s functions by DISCOVER and learned Green’s functions by neural network (black box models). The first five eigenfunctions of (b) the exact Green’s functions, (c) learned Green’s functions by DISCOVER, (d) learned Green’s functions by neural network (black box models).}
    \label{fig6}
\end{figure}

To further validate our findings, we also perform an eigenvalue decomposition of the Green's function in this case. As depicted in FIG. \ref{fig6}, the eigenvalues and dominant eigenmodes are accurately recovered, although there are minor differences in the coefficients of the Green's function discovered by DISCOVER compared to the exact Green's function. Specifically, FIG. \ref{fig6} (a) illustrates that despite these minor coefficient discrepancies, DISCOVER effectively recovers both the eigenvalues and the dominant eigenmodes. It is noteworthy that, in this case, while neural networks achieve performance comparable to the DISCOVER framework, as shown in FIG. \ref{fig3}, they still exhibit a decay in the eigenvalues of the learned Green's function. The eigenvalues of the Green's function approximated by the rational neural network align with the actual eigenvalues only for the first 25 values. In contrast, the eigenvalues of the Green's function identified by DISCOVER match the true eigenvalues entirely. 
This observation may be due to the neural network's smooth approximation of the exact Green's function \cite{boulle2022data}. Therefore, the DISCOVER framework is deemed more advantageous for scientific discovery, as it provides deeper insights into the physical core information.

\subsection{Discovery results of unknown Green's functions.}
\subsubsection*{Case 3: Periodic helmholtz operator.}
Periodic boundary conditions play a pivotal role in fluid dynamics, enabling the simulation of infinite or semi-infinite domains by replicating a fundamental cell's behavior. This approach is particularly instrumental in computational fluid dynamics, where it facilitates the study of flow phenomena over extended regions without the computational overhead of large-scale simulations. Moreover, periodic boundary conditions are essential for analyzing wave-like behavior and turbulence in fluid systems, providing insights into their inherent cyclical patterns. In this example, we aim to learn the unknown Green’s function of a Helmholtz operator with periodic boundary conditions on $\Omega=[0,1]$ with frequency $K=3$:
\begin{equation}
\left\{
\begin{aligned}
    \mathcal{L} u &=\frac{d^2 u}{d x^2}+K^2 u=f, \quad &x \in [0,1],\\
    u(0)&=u(1).
\end{aligned}
\right.
\end{equation}

The Green's function of this case is unknown. We only use 100 input-output pairs $\left\{\left(f_j, u_j\right)\right\}_{j=1}^N$ calculated by spectral method as stated in section \ref{sec2}. The expression extracted by DISCOVER is:
\begin{equation}
\mathcal{T}(x, y)=\left\{
\begin{aligned}
    \mathcal{T}_L(x, y) &= -0.167085 \cdot sin[\mathbf{3}\left(-0.023598+x-y\right)], & \text { if } x \leq y,\\
    \mathcal{T}_R(x, y) &= -0.167085 \cdot sin[\mathbf{3}\left(-0.023598+y-x\right)]. & \text { if } x>y,
\end{aligned}
\right.
\label{eq22}
\end{equation}

\begin{figure}[H]
    \centering
    \includegraphics[width=0.81\textwidth]{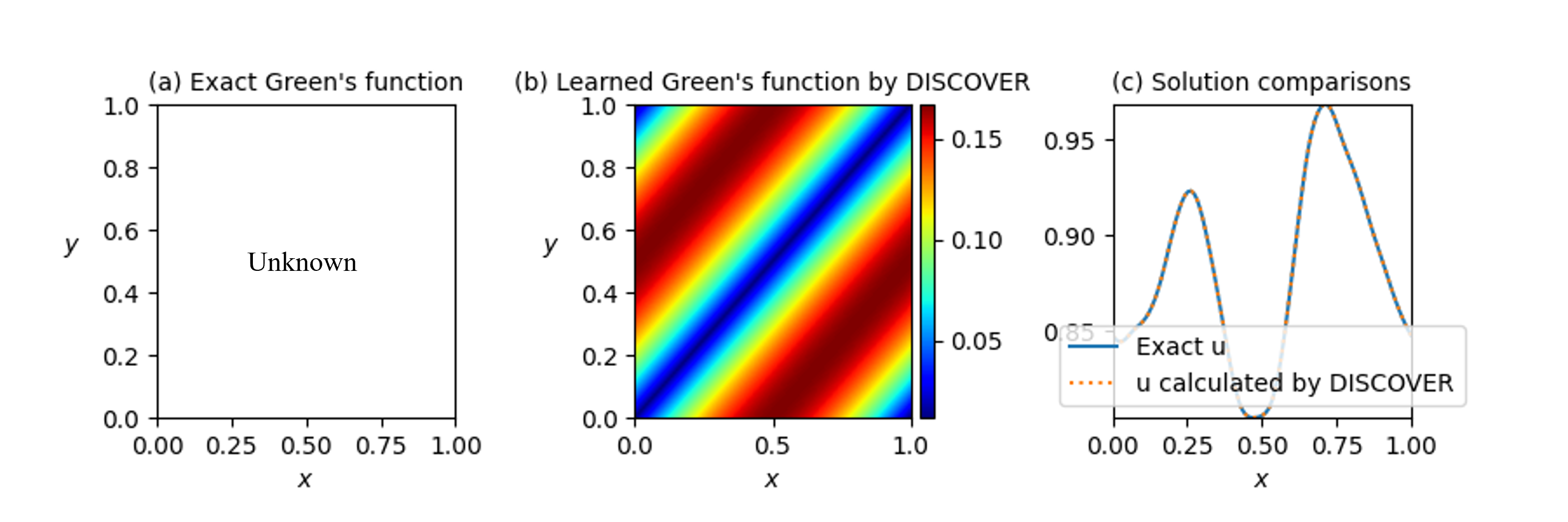}
    \caption{Periodic Helmholtz operators for $K=3$. (a) Exact Green's function. (b) Learned Green’s functions by DISCOVER. (c) Exact solution and solution calculated by learned Green's function.}
    \label{fig4}
\end{figure}

The reward of the learned Green's function calculated by Eq. (\ref{eq9}) is 0.9899, whilst the MSE error calculated by Eq. (\ref{eq10}) is $6.2547\times 10^{-11}$. The learned Green’s function and calculated solution is displayed in FIG. \ref{fig4}. 

In FIG. \ref{fig4}, it is evident that the Green's function itself exhibits periodicity, as indicated by $G(0, y)=G(1, y)$ for all $y \in [0,1]$, in line with expectations. This periodicity in the $y$-direction, expressed as $G(x, 0)=G(x, 1)$ for $x \in [0,1]$, arises from the self-adjoint nature of the Helmholtz operator, implying symmetry in the associated Green's function. 
Moreover, any linear constraint $\mathcal{C}(u)=0$, such as linear conservation laws or symmetries \cite{olver1993applications}, that holds true for all solutions to Eq. (\ref{eq1}) under forcing $f \in C_c^{\infty}(\Omega)$ is also satisfied by the Green's function, $G$. This is evidenced by the condition $\mathcal{C}(G(\cdot, y))=0$ for all $y \in \Omega$, thereby affirming the observance of such constraints by the Green's function.

It is imperative to acknowledge the significance of the symmetry hard constraint in the derivation of the Green's function. In the absence of this constraint, the resulting expression, exemplified by \(\sin(\sin(x+1.13-y(x>y))-7.14)\), yields a reward of 0.6943 and an MSE of \(1.30 \times 10^{-2}\). However, when the symmetry hard constraint is enforced, the solution's accuracy can be elevated to an order of \(10^{-11}\). This level of precision surpasses that attainable by conventional approximation techniques, including those employed by neural networks. Consequently, the necessity of the symmetry hard constraint in the discovery process of the Green's function is underscored.

\begin{figure}[H]
    \centering
    \includegraphics[width=0.81\textwidth]{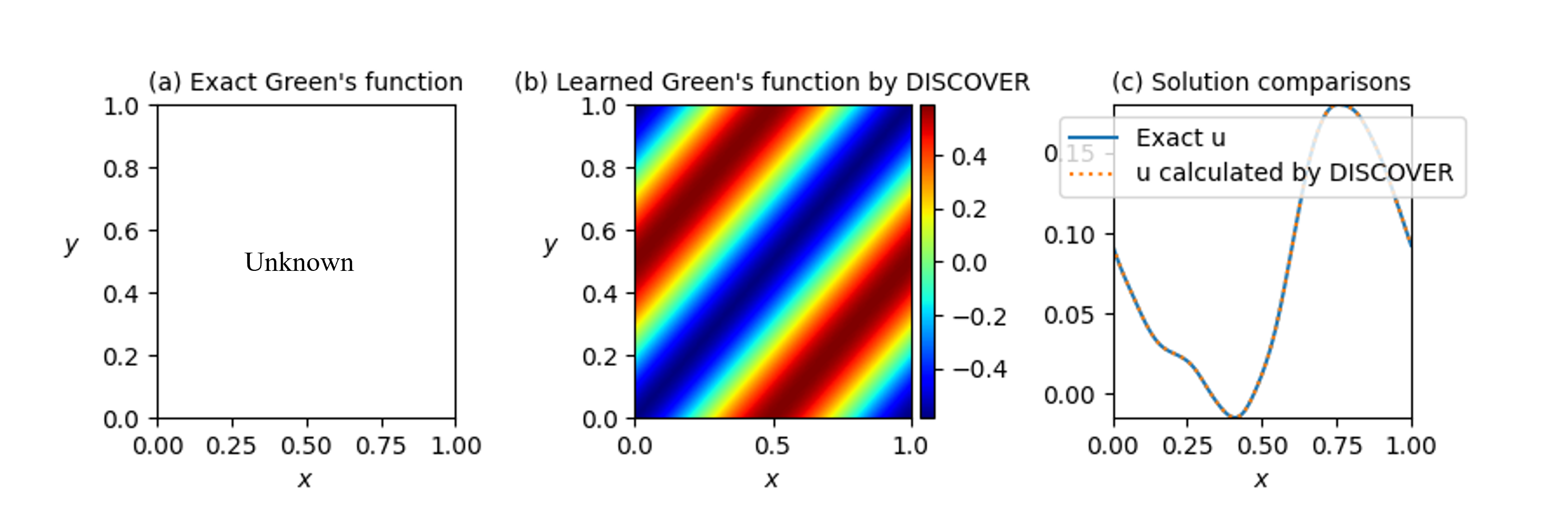}
    \caption{Periodic Helmholtz operators for $K=6$. (a) Exact Green's function. (b) Learned Green’s functions by DISCOVER. (c) Exact solution and solution calculated by learned Green's function.}
    \label{fig8}
\end{figure}

\begin{figure}[H]
    \centering
    \includegraphics[width=0.81\textwidth]{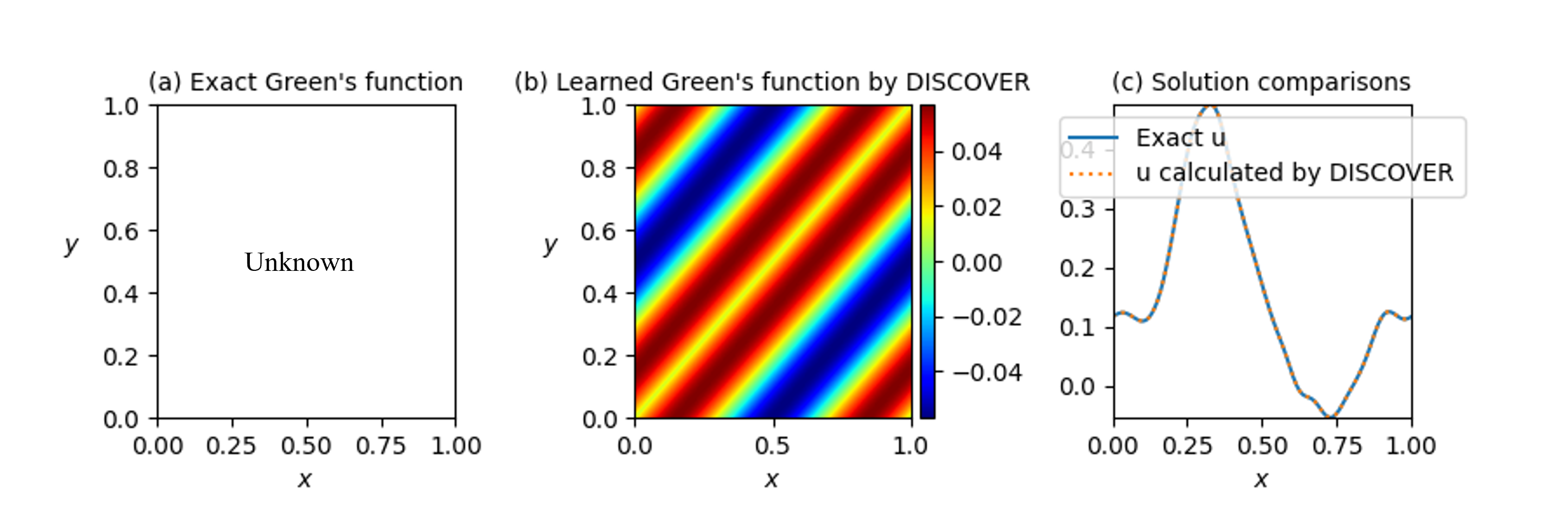}
    \caption{Periodic Helmholtz operators for $K=9$. (a) Exact Green's function. (b) Learned Green’s functions by DISCOVER. (c) Exact solution and solution calculated by learned Green's function.}
    \label{fig9}
\end{figure}

Furthermore, it is intriguing to observe that the coefficient 3 emerges in Eq. (\ref{eq22}), coinciding precisely with the value of the frequency $K=3$. To investigate whether this alignment is merely fortuitous or indicative of a deeper pattern, we undertook an extensive search for corresponding expressions across a spectrum of $K$ values. The outcomes of this exploration are meticulously documented in Table \ref{table4}. The Green's functions derived using the DISCOVER method exhibit remarkably low MSE, on the order of \(10^{-10}\). Such a level of precision is challenging to attain with approximation methods, including neural networks. Consequently, these Green's functions can be considered accurate. This result further substantiates the superiority of the DISCOVER method in identifying unknown equations. It is noteworthy that all discovered Green's functions exhibit a uniform expression structure, with variations solely in their coefficients. We display the results of $K=6$ and $K=9$ in Figures \ref{fig8}-\ref{fig9}, where the shapes of the Green's function are similar and are relevant with the frequency $K$. Additionally, it has proven challenging to identify a qualified Green's function for cases where $K$ exceeds 10. This difficulty may stem from the fact that the coefficient associated with the $\sin$ term becomes exceedingly small due to the high frequency of $K$, thereby complicating the discernment of a discernible pattern or function. This phenomenon underscores the complexity inherent in the analysis of Green's functions at higher frequencies and suggests the need for further research to elucidate the underlying mechanisms governing their behavior.

\begin{table}[H]
\centering
\caption{Summary of discovered results for different $K$.}
\label{table4}
\begin{tabular}{clcc}
\toprule
 $K$ & \multicolumn{1}{c}{DISCOVERed Green's function} & Reward & MSE \\
 \hline
$\mathbf{1}$ & $\quad\mathcal{T}(x, y)=\left\{
\begin{aligned}
    \mathcal{T}_L(x, y) &= 1.042918 \cdot \sin[\mathbf{1}\left(2.070786 + x - y\right)], & \text{if } x \leq y, \\
    \mathcal{T}_R(x, y) &= 1.042918 \cdot \sin[\mathbf{1}\left(2.070786 + y - x\right)], & \text{if } x > y. 
\end{aligned}
\right.$ & 0.9899 & $6.6391 \times 10^{-12}$ \\
$\mathbf{2}$ & $\quad\mathcal{T}(x, y)=\left\{
\begin{aligned}
    \mathcal{T}_L(x, y) &= 0.297098 \cdot \sin[\mathbf{2}\left(0.904866+x-y\right)], & \text { if } x \leq y,\\
    \mathcal{T}_R(x, y) &= 0.297098 \cdot \sin[\mathbf{2}\left(0.904866+y-x\right)]. & \text { if } x>y.
\end{aligned}
\right.$ & 0.9899 & $2.6219 \times 10^{-11}$ \\
$\mathbf{4}$ & $\quad\mathcal{T}(x, y)=\left\{
\begin{aligned}
    \mathcal{T}_L(x, y) &= -0.137468 \cdot \sin[\mathbf{4}\left(0.107301+x-y\right)], & \text { if } x \leq y,\\
    \mathcal{T}_R(x, y) &= -0.137468 \cdot \sin[\mathbf{4}\left(0.107301+y-x\right)]. & \text { if } x>y.
\end{aligned}
\right.$ & 0.9899 & $1.5890 \times 10^{-10}$ \\
$\mathbf{5}$ & $\quad\mathcal{T}(x, y)=\left\{
\begin{aligned}
    \mathcal{T}_L(x, y) &= -0.167092 \cdot \sin[\mathbf{5}\left(0.185842+x-y\right)], & \text { if } x \leq y,\\
    \mathcal{T}_R(x, y) &= -0.167092 \cdot \sin[\mathbf{5}\left(0.185842+y-x\right)]. & \text { if } x>y.
\end{aligned}
\right.
$ & 0.9899 & $8.6256 \times 10^{-11}$ \\
$\mathbf{6}$ & $\quad\mathcal{T}(x, y)=\left\{
\begin{aligned}
    \mathcal{T}_L(x, y) &= -0.590514 \cdot \sin[\mathbf{6}\left(0.2382+x-y\right)], & \text { if } x \leq y,\\
    \mathcal{T}_R(x, y) &= -0.590514 \cdot \sin[\mathbf{6}\left(0.2382+y-x\right)]. & \text { if } x>y.
\end{aligned}
\right.
$ & 0.9899 & $7.3422 \times 10^{-12}$ \\
$\mathbf{7}$ & $\quad\mathcal{T}(x, y)=\left\{
\begin{aligned}
    \mathcal{T}_L(x, y) &= 0.203626 \cdot \sin[\mathbf{7}\left(0.2756+x-y\right)], & \text { if } x \leq y,\\
    \mathcal{T}_R(x, y) &= 0.203626 \cdot \sin[\mathbf{7}\left(0.2756+y-x\right)]. & \text { if } x>y.
\end{aligned}
\right.
$ & 0.9899 & $6.3973 \times 10^{-11}$ \\
$\mathbf{8}$ & $\quad\mathcal{T}(x, y)=\left\{
\begin{aligned}
    \mathcal{T}_L(x, y) &= 0.082585 \cdot \sin[\mathbf{8}\left(0.303651+x-y\right)], & \text { if } x \leq y,\\
    \mathcal{T}_R(x, y) &= 0.082585 \cdot \sin[\mathbf{8}\left(0.303651+y-x\right)]. & \text { if } x>y.
\end{aligned}
\right.
$ & 0.9899 & $2.6441 \times 10^{-10}$ \\
$\mathbf{9}$ & $\quad\mathcal{T}(x, y)=\left\{
\begin{aligned}
    \mathcal{T}_L(x, y) &= 0.056832 \cdot \sin[\mathbf{9}\left(0.325468+x-y\right)], & \text { if } x \leq y,\\
    \mathcal{T}_R(x, y) &= 0.056832 \cdot \sin[\mathbf{9}\left(0.325468+y-x\right)]. & \text { if } x>y.
\end{aligned}
\right.
$ & 0.9899 & $7.5148 \times 10^{-10}$ \\
$\mathbf{10}$ & $\quad\mathcal{T}(x, y)=\left\{
\begin{aligned}
    \mathcal{T}_L(x, y) &= 0.052142 \cdot \sin[\mathbf{10}\left(0.342921+x-y\right)], & \text { if } x \leq y,\\
    \mathcal{T}_R(x, y) &= 0.052142 \cdot \sin[\mathbf{10}\left(0.342921+y-x\right)]. & \text { if } x>y.
\end{aligned}
\right.
$ & 0.9899 & $4.4224 \times 10^{-10}$ \\
\bottomrule
\end{tabular}
\end{table}

\subsubsection*{Case 4: Second-order differential operator with jump conditions.}
The solution of second-order differential operators with jump conditions is of paramount importance across various scientific and engineering disciplines. In materials science and engineering, it facilitates the modeling of discontinuities in heterogeneous materials and the analysis of fracture mechanics. Fluid dynamics benefits from accurately describing multiphase flows and shock waves, while electromagnetics leverages these solutions for interface problems involving different media. In geophysics, it aids in the precise modeling of seismic wave propagation through the Earth's stratified layers. In biological systems, it is used to model discontinuities in stress or strain at tissue interfaces, and in financial mathematics, it is essential for pricing derivatives in jump-diffusion models of asset pricing. These applications underscore the critical role of solving differential equations with jump conditions in enhancing the accuracy of simulations and predictions in complex systems.

In this case, we display the discovery results for the Green's function of a second-order differential operator with a jump condition, defined on $\Omega=[0,1]$ as
\begin{equation}
    \left\{
\begin{aligned}
    &\mathcal{L} u =0.2 \frac{d^2 u}{d x^2}+\frac{d u}{d x}=f, \quad x \in [0,1],\\
    &u(0) =0, \quad u(1)=0, \\
    &u(0.7^{-}) =2, \quad u(0.7^{+})=1.
\end{aligned}
\right.
\end{equation}

The expression extracted by DISCOVER is:
\begin{equation}
\mathcal{T}(x, y)=\left\{
\begin{aligned}
    \mathcal{T}_L(x, y) &= x \cdot \left( 0.128885-e^{2.244143-4.429130 y} \right), & \text { if } x \leq y,\\
    \mathcal{T}_R(x, y) &= y \cdot \left( 0.128885-e^{2.244143-4.429130 x} \right). & \text { if } x>y.
\end{aligned}
\right.
\end{equation}

The performance metrics for the learned Green's function are impressive: the reward calculated by Eq. (\ref{eq9}) is 0.9564, indicating a high level of fidelity to observed data. Moreover, the MSE, computed using Eq. (\ref{eq10}), is $6.0825\times 10^{-4}$, underscoring the accuracy of the approximation.

\begin{figure}[H]
    \centering
    \includegraphics[width=0.95\textwidth]{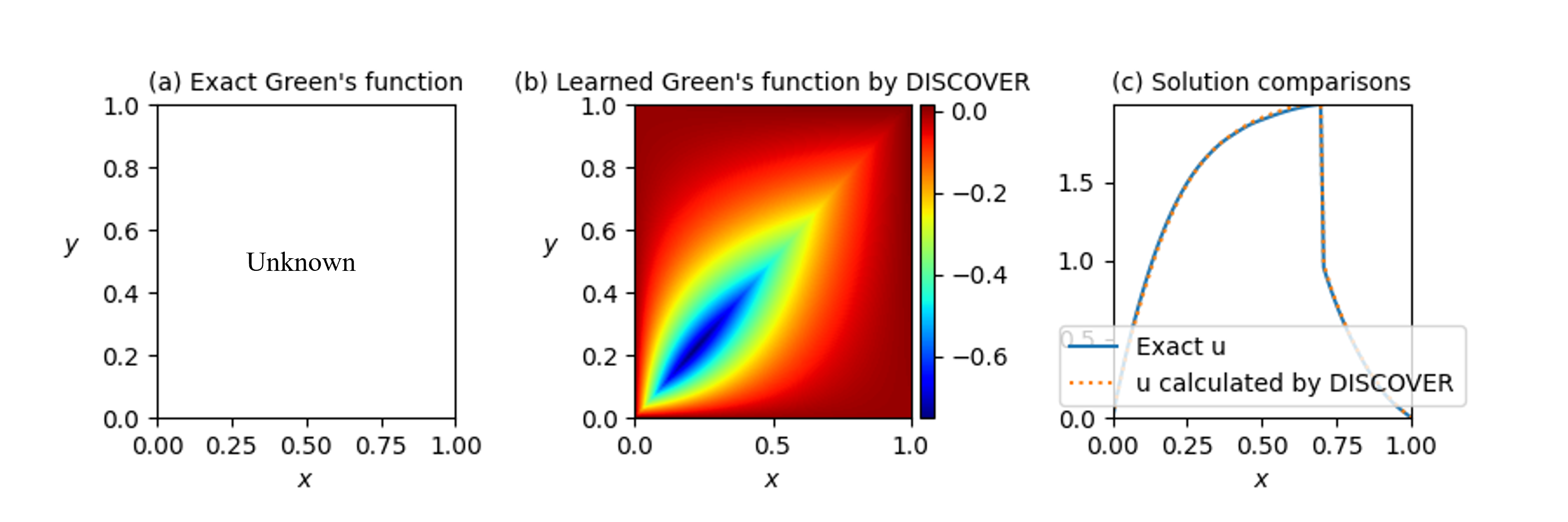}
    \caption{Second-order differential operator with a jump conditions $x=0.7$. (a) Exact Green’s function. (b) Learned Green’s functions by DISCOVER. (c) Exact solution and solution calculated by learned Green’s function.}
    \label{fig5}
\end{figure}

The learned Green’s function and the corresponding calculated solution are displayed in FIG. \ref{fig5}. The results underscore the efficacy of the DISCOVER framework in clustering poles, which significantly aids in the identification of their precise locations and types. A meticulous examination of this clustering provides insightful details about the nature of the singularities inherent in the Green's function. Specifically, the singularity observed at \(x = 0.7\) is indicative of a jump condition. This observation highlights the DISCOVER framework's capability to not only learn and approximate Green's functions with high precision but also to offer a deeper understanding of the underlying physical phenomena. By identifying discontinuities and irregularities within the solution domain, the DISCOVER framework proves especially valuable for analyzing complex differential operators. Traditional methods often struggle to accurately capture and interpret such features, making DISCOVER a superior tool in these contexts. Furthermore, the ability to pinpoint and characterize singularities is crucial for advancing our comprehension of the physical processes modeled by these differential operators, thereby facilitating more accurate and insightful analyses.

\subsection{Noise perturbation.\label{noise_sec}}
Since observations obtained from real scenes are often noisy, it is necessary to test whether DISCOVER can mine the correct Green's function under different noise levels. We perturb the system’s response measurements with Gaussian noise as:
\begin{equation}
    u_j^{\text {noise}}\left(x_i\right)=u_j\left(x_i\right)\left(1+\delta c_{i, j}\right)
\end{equation}
where the coefficients $c_{i, j}$ are independent and identically distributed, Gaussian random variables for $1 \leq i \leq N_u$ and $1 \leq j \leq N$, and $\delta$ denotes the noise level (in percent). We then vary the level of Gaussian noise perturbation from $0 \%$ to $10 \%$, train the DISCOVER for each choice of the noise level, and report the relative error in Table \ref{table2}. 

As shown in Table \ref{table2}, the robustness of the DISCOVER algorithm under various noise levels is exemplified through a series of tests. Across different cases, the algorithm demonstrates a consistent performance with minimal degradation in accuracy and MSE, even when subjected to significant noise perturbations.

For instance, in the first case with an exact Green's function, the algorithm achieves a high reward of 0.9899 and an MSE of \(9.185 \times 10^{-10}\) under 1\% noise. This performance remains robust with an MSE of \(2.79 \times 10^{-9}\) and a slightly reduced reward of 0.9897 under 5\% noise. Even at a 10\% noise level, the reward is still high at 0.9891 with an MSE of \(2.83 \times 10^{-8}\), indicating a remarkable stability in the face of noise.

In the second case, involving a more complex Green's function with hyperbolic sine functions, the algorithm shows a slight decrease in reward from 0.9891 at 0\% noise to 0.8456 at 5\% noise, yet the MSE remains within an acceptable range, increasing from \(5.10 \times 10^{-8}\) to \(3.39 \times 10^{-3}\). Moreover, there exists high-frequency characteristics specific to case 2, where the exact Green's function exhibits an exceptionally small coefficient on the order of \(10^{-5}\). As a result, accurately capturing the modes of the Green's function becomes challenging. This suggests that while the algorithm's precision is affected by higher noise levels, it still maintains a relatively high level of performance.

In the third and fourth cases, where the exact Green's functions are unknown, the algorithm continues to exhibit commendable robustness. For instance, in the third case, the reward remains above 0.98 even at 10\% noise, with an MSE of \(7.94 \times 10^{-8}\). Similarly, in the fourth case, the algorithm's performance is only slightly impacted by noise, with the reward dropping from 0.9457 at 1\% noise to 0.9478 at 10\% noise, and the MSE increasing from \(1.08 \times 10^{-3}\) to \(9.74 \times 10^{4}\).

These results underscore the DISCOVER algorithm's resilience and reliability in noisy environments, a critical attribute for applications where data integrity may be compromised by external factors. The capacity to handle noise effectively while capturing intricate features of Green's functions highlights the strength of the DISCOVER approach, making it a valuable asset in the toolkit for solving differential equations in both theoretical research and practical engineering applications.

\begin{table}
\centering
\caption{Summary of discovered results for different operators perturbing by various noise levels.}
\label{table2}
\begin{threeparttable}
\begin{tabular}{cccc}
\toprule
\multicolumn{4}{c}{\textbf{Case 1:} Exact Green's function: $(x-xy)\cdot(x<=y)+(y-xy)\cdot(x>y)$}\\
\hline
 Noise & Discovered Green's function & Reward & MSE \\
 0 & $(1.00x-1.00xy)\cdot(x<=y)+(1.00y-1.00xy)\cdot(x>y)$ & \textbf{0.9899} & $1.48\times10^{-10}$\\
 1\% & $x(1.00-1.00y+1.26\times10^{-5}x)\cdot(x<=y)+y(1.00-1.00x+1.26\times10^{-5}y)\cdot(x>y)$ & 0.9899 & $1.85\times10^{-10}$ \\
 5\% & $(x+2.41\times10^{-8})(1.00-1.00y)\cdot(x<=y)+(y+2.41\times10^{-8})(1.00-1.00x)\cdot(x>y)$ & 0.9897 & $2.79\times10^{-9}$ \\
 10\% & $(1.00x-1.00xy-4.99\times10^{-9})\cdot(x<=y)+(1.00y-1.00xy-4.99\times10^{-9})\cdot(x>y)$ & 0.9891 & $2.83\times10^{-8}$ \\
\hline
\multicolumn{4}{c}{\textbf{Case 2:} Exact Green's function: $\frac{\sinh(8x)\sinh(8(y-1))}{8\sinh(8)}\cdot(x<=y)+\frac{\sinh(8y)\sinh(8(x-1))}{8\sinh(8)}\cdot(x>y)$}\\
\hline
 Noise & Discovered Green's function & Reward & MSE \\
 0 & $8.27\times10^{-5}[\sinh(8.02x)\sinh(8.02y-8.01)(x<=y)+\sinh(8.02y)\sinh(8.02x-8.01)(x>y)]$ & \textbf{0.9891} & $5.10\times 10^{-8}$\\
 1\%  & $8.27\times10^{-5}[\sinh(7.99x)\sinh(7.99y-7.99)(x<=y)+\sinh(7.99y)\sinh(7.99x-7.99)(x>y)]$ & 0.9800 & $5.05\times 10^{-8}$  \\
 5\% & $4.75\times10^{-6}[\sinh(x)\sinh(-12.26-1.49x+8.28y)(x<=y)+\sinh(y)\sinh(-12.26-1.49y+8.28x)(x>y)]$ & 0.8597 & $2.59\times10^{-3}$ \\
 10\% & $3.735\times10^{-3}[\sinh(x)\sinh(-5.74-5.3652x+8.2y)(x<=y)+\sinh(y)\sinh(-5.74-5.3652y+8.2x)(x>y)]$ & 0.8456 & $2.78\times10^{-3}$ \\
\hline
\multicolumn{4}{c}{\textbf{Case 3:} Exact Green's function: Unknown}\\
\hline
 Noise & Discovered Green's function & Reward & MSE \\
 0 & $0.17\sin(3(0.024-x+y))(x<=y)+0.17\sin(3(0.024-y+x))(x>y)]$ & \textbf{0.9899} & $6.25\times 10^{-11}$\\
 1\% & $0.17\sin(0.07+3y-3x)(x<=y)+0.17\sin(0.07-3y+3x)(x>y)]$ & 0.9892 & $3.72\times 10^{-8}$ \\
 5\% & $(-1.14-1.3\sin(y-x+4.21))\cdot(x<=y)+(-1.14-1.3\sin(x-y+4.21))\cdot(x>y)$ & 0.9731 & $2.16\times10^{-5}$ \\
 10\% & $0.17\sin(3.06-2.98(x-y))\cdot(x<=y)+0.17\sin(3.06-2.98(y-x))\cdot(x>y)$ & 0.9889 &  $7.94\times10^{-8}$\\
\hline
\multicolumn{4}{c}{\textbf{Case 4:} Exact Green's function: Unknown}\\
\hline
 Noise & Discovered Green's function & Reward & MSE \\
 0 & $x(0.13-e^{2.24-4.43y})(x<=y)+y(0.13-e^{2.24-4.43x})(x>y)$ & \textbf{0.9564} & $6.08\times 10^{-4}$\\
 1\%  & $((0.90-y)xe^x-3.76)\cdot(x<=y)+((0.90-x)ye^y-3.76)\cdot(x>y)$ & 0.9457 & $1.08\times10^{-3}$ \\
 5\%  & $-4.27xe^{-5.93y^2}\cdot(x<=y)-4.27ye^{-5.93x^2}\cdot(x>y)$ & 0.9593 & $5.05\times10^{-4}$ \\
 10\%  & $-3.17ye^{x-2.83(y+y^2)}\cdot(x<=y)-3.17xe^{y-2.83(x+x^2)}\cdot(x>y)$ & 0.9478 & $9.74\times10^{-4}$ \\
\bottomrule
\end{tabular}
\end{threeparttable}
\end{table}

\subsection{Ablation Studies.\label{ablation_sec}}
Ablation experiments are a research method used to systematically remove or disable certain components of a system or model to evaluate their individual contributions and overall importance to the system's performance. In this section, we conduct specific ablation studies to validate the effectiveness of the symmetric constraint in learning Green's function.
Specifically, to rigorously assess the impact of this term on the learning outcomes, we conduct ablation experiments in two distinct configurations: one that integrates symmetric constraints and one that omits them entirely. The comparative evaluation, as detailed in Table \ref{table3}, reveals a significant enhancement in the accuracy and reliability of the discovered Green's function when symmetric constraints are incorporated.

Table \ref{table3} provides a comparative analysis of the discovered Green's functions under different constraint scenarios, highlighting the significant impact of these constraints on both the reward and MSE.

For instance, in Case 1, the absence of symmetry constraints (\texttimes) leads to a Green's function that yields a reward of 0.6608 and an MSE of \(8.53 \times 10^3\), which is notably inferior to the performance achieved with symmetry constraints (\checkmark), where the reward is 0.9899 and the MSE is as low as \(1.48 \times 10^{-10}\). This stark contrast highlights the effectiveness of symmetry constraints in refining the algorithm's output.

Similarly, in Case 2, the introduction of symmetry constraints (\checkmark) results in a reward of 0.9891 and an MSE of \(5.10 \times 10^{-5}\), whereas without these constraints (\texttimes), the reward plummets to 0.6037 and the MSE escalates to 0.026. This further substantiates the hypothesis that symmetry constraints are instrumental in optimizing the algorithm's performance.

In Case 3, the impact of symmetry constraints is again evident. With constraints (\checkmark), the algorithm achieves a reward of 0.9899 and an MSE of \(6.25 \times 10^{-11}\), whereas without constraints (\texttimes), the reward decreases to 0.6943 and the MSE increases to \(1.30 \times 10^{-3}\). This disparity underscores the constraints' contribution to the accuracy and precision of the discovered Green's functions.

Lastly, in Case 4, the presence of symmetry constraints (\checkmark) secures a reward of 0.9564 and an MSE of \(6.08 \times 10^{-4}\), while their absence (\texttimes) results in a slightly lower reward of 0.9387 and a higher MSE of \(1.47 \times 10^{-3}\). Although the difference here is not as pronounced as in the previous cases, it still indicates the beneficial effect of symmetry constraints on the algorithm's performance.

In conclusion, the ablation study provides compelling evidence that symmetry constraints are crucial for the DISCOVER algorithm's ability to accurately and efficiently discover Green's functions. The consistent improvement in reward and reduction in MSE with the application of these constraints validate their significance in the algorithmic process. This advancement demonstrates the critical importance of embedding theoretical constraints into machine learning frameworks to align computational models more closely with underlying physical laws. Consequently, the improved DISCOVER framework stands as a powerful tool for solving complex differential equations, offering significant benefits for research and practical applications in diverse fields such as physics, engineering, and applied mathematics.

\begin{table}[H]
\centering
\caption{Ablation studies of discovered results for different operators.}
\label{table3}
\begin{tabular}{cccc}
\toprule
\multicolumn{4}{c}{\textbf{Case 1:} Exact Green's function: $(x-xy)\cdot(x<=y)+(y-xy)\cdot(x>y)$}\\
\hline
Constraints & Discovered Green's function & Reward & MSE \\
 \checkmark & $(1.00x-1.00xy)\cdot(x<=y)+(1.00y-1.00xy)\cdot(x>y)$ & \textbf{0.9899} & $1.48\times10^{-10}$\\
 \texttimes & $[0.51-0.26(x>y)-0.51x]\cdot (y-0.002)$ & 0.6608 & $8.53\times10^{-3}$ \\
\hline
\multicolumn{4}{c}{\textbf{Case 2:} Exact Green's function: $\frac{\sinh(8x)\sinh(8(y-1))}{8\sinh(8)}\cdot(x<=y)+\frac{\sinh(8y)\sinh(8(x-1))}{8\sinh(8)}\cdot(x>y)$}\\
\hline
 Constraints & Discovered Green's function & Reward & MSE \\
 \checkmark & $8.27\times10^{-5}[\sinh(8.02x)\sinh(8.02y-8.01)(x<=y)+\sinh(8.02y)\sinh(8.02x-8.01)(x>y)]$ & \textbf{0.9891} & $5.10\times 10^{-8}$\\
 \texttimes & $0.52\sinh((y-x)(y-\sinh(x)+0.03))-0.018$ & 0.6037 & 0.026 \\
\hline
\multicolumn{4}{c}{\textbf{Case 3:} Exact Green's function: Unknown}\\
\hline
 Constraints & Discovered Green's function & Reward & MSE \\
 \checkmark & $0.17\sin(3(0.90-x+y))(x<=y)+0.17\sin(3(0.90-y+x))(x>y)]$ & \textbf{0.9899} & $6.25\times 10^{-11}$\\
 \texttimes & $\sin(\sin(x+1.13-y(x>y))-7.14)$ & 0.6943 & $1.30\times10^{-2}$ \\
\hline
 \multicolumn{4}{c}{\textbf{Case 4:} Exact Green's function: Unknown}\\
\hline
 Constraints & Discovered Green's function & Reward & MSE \\
 \checkmark & $x(0.13-e^{2.24-4.43y})(x<=y)+y(0.13-e^{2.24-4.43x})(x>y)$ & \textbf{0.9564} & $6.08\times 10^{-4}$\\
 \texttimes & $(0.86-y)(y(e^{x+0.48}-3.7)-0.07)$ & 0.9387 & $1.47\times10^{-3}$ \\
\bottomrule
\end{tabular}
\end{table}

\section{Conclusions}
In this paper, we utilize the DISCOVER framework with physical hard constraints to find unknown Green's functions. Leveraging these discovered Green's functions, we can ascertain analytical solutions for specified PDEs, thereby enabling efficient problem-solving and providing a systematic understanding of the governing theorems. 
The application of symbolic regression to uncover Green’s functions represents a significant milestone in employing artificial intelligence to accelerate scientific discovery.

We specifically identified the Green's functions for the Laplace and Helmholtz operators, which exactly match the ground-truth Green’s functions. Additionally, we compared our results with those obtained using PySR and rational neural networks. The accuracy of DISCOVER surpasses that of PySR, a commonly used method in the symbolic regression field, and even achieves the same level of accuracy as black-box neural network models while utilizing significantly fewer parameters. More importantly, compared to neural networks that can only approximate the distribution of the Green's function without accurately capturing its eigenvalues, DISCOVER excels at achieving both. This highlights DISCOVER's superior capability for scientific discovery, as it provides more fundamental physical insights.

Furthermore, we discovered the Green's functions for the periodic Helmholtz operator and second-order differential operator with jump conditions, where exact Green's functions are unknown. For the periodic Helmholtz operator, we achieved a solution error on the order of $10^{-10}$, a level of precision challenging to attain with traditional approximation methods like neural networks. This result further substantiates the superiority of the DISCOVER method in identifying unknown equations.

The Laplace and Helmholtz operators play a crucial role in solving the pressure Poisson equation, whereas second-order differential operators with jump conditions are indispensable for the study of multiphase flows and shock waves. This utilization of symbolic regression for uncovering Green's functions marks a significant breakthrough in employing artificial intelligence to expedite scientific discoveries, particularly in fluid dynamics and associated disciplines. Future research will aim to expand this framework to identify unknown Green's functions in two-dimensional and three-dimensional scenarios.

\section*{Acknowledgements}
This work was supported by the National Natural Science Foundation of China (Grant No. 62106116),  China  Meteorological  Administration  under  Grant  QBZ202316,  Natural  Science Foundation of Ningbo of China (No. 2023J027), as well as by the High Performance Computing Centers at Eastern Institute of Technology, Ningbo, and  Ningbo Institute of Digital Twin.

\section*{Reference}
\bibliography{aipsamp}

\end{document}